# Nanoscale imaging of equilibrium quantum Hall edge currents and of the magnetic monopole response in graphene

Aviram Uri[1*], Youngwook Kim[2,3], Kousik Bagani[1], Cyprian K. Lewandowski[4], Sameer Grover[1], Nadav Auerbach[1], Ella O. Lachman[1†], Yuri Myasoedov[1], Takashi Taniguchi[5], Kenji Watanabe[5], Jurgen Smet[2], and Eli Zeldov[1*]

[1]*Department of Condensed Matter Physics, Weizmann Institute of Science, Rehovot 7610001, Israel*

[2]*Max Planck Institute for Solid State Research, D-70569 Stuttgart, Germany*

[3]*Department of Emerging Materials Science, DGIST, Dalseong-Gun, Daegu, 42988 Korea*

[4]*Department of Physics, Massachusetts Institute of Technology, Cambridge, MA 02139, USA*

[5]*National Institute for Material Science, 1-1 Namiki, Tsukuba, 305-0044, Japan*

[†]*Current address: Department of Physics, University of California, Berkeley, CA 94720, USA*

[*]*Corresponding authors*

**Abstract**

The recently predicted topological magnetoelectric effect [1] and the response to an electric charge that mimics an induced mirror magnetic monopole [2] are fundamental attributes of topological states of matter with broken time reversal symmetry. Using a SQUID-on-tip [3], acting simultaneously as a tunable scanning electric charge and as ultrasensitive nanoscale magnetometer, we induce and directly image the microscopic currents generating the magnetic monopole response in a graphene quantum Hall electron system. We find a rich and complex nonlinear behavior governed by coexistence of topological and nontopological equilibrium currents that is not captured by the monopole models [2]. Furthermore, by utilizing a tuning fork that induces nanoscale vibrations of the SQUID-on-tip, we directly image the equilibrium currents of individual quantum Hall edge states for the first time. We reveal that the edge states that are commonly assumed to carry only a chiral downstream current, in fact carry a pair of counterpropagating currents [4], in which the topological downstream current in the incompressible region is always counterbalanced by heretofore unobserved nontopological upstream current flowing in the adjacent compressible region. The intricate patterns of the counterpropagating equilibrium-state orbital currents provide new insights into the microscopic origins of the topological and nontopological charge and energy flow in quantum Hall systems.

Magnetic monopole is a hypothetical elementary particle representing an isolated source of magnetic field with only one magnetic pole (N without S, or vice versa) and a quantized magnetic charge. While modern particle theory predicts its existence, experimental searches for magnetic monopoles have so far been unsuccessful. Condensed matter systems, however, offer a natural platform for studying magnetic-monopole-like excitations or response in materials like spin ice [5,6] or topological insulators owing to their topological magnetoelectric effect (TME) [1,7,8]. The conventional, nontopological magnetoelectric effect (ME) describes magnetization induced by externally applied electric field in materials in which the time-reversal symmetry is broken either by magnetic order or by an applied magnetic field [9]. The conventional ME is distinctly local in nature – the local magnetization $\boldsymbol{m}$ is proportional to the local electric field $\boldsymbol{E}$ through $m_i = \alpha_{ij} E_j$, where $\alpha_{ij}$ is the so-called Lifshitz tensor [9]. In the last decades this ME has enabled key advances in spin-based information processing and has been exploited in fast-access magnetic memory of keen practical interest [10].

The advent of topology has led to the notion that topological systems can exhibit a magnetoelectric effect of an entirely different nature with two fundamentally different characteristics – nonlocality and topologically quantized response [1,7,8]. This TME is predicted to arise in three-dimensional magnetic topological insulators and in two-dimensional electron systems with quantized Hall conductivity. In these systems topological currents may flow in the ground state. They are expected to generate an exotic response in the form of a mirror magnetic monopole characterized by the Chern number $C$ [2,11,12]. One can depict this TME by considering a point charge $Q_e$ placed at $r = 0$ and height $z_0$ above the topological surface (Fig. 1a). In this case, the induced circulating topological surface current density

$$J_\varphi^T(r) = \frac{r\hat{\varphi}}{2\pi(r^2+z_0^2)^{3/2}} Q'_m \qquad (1)$$

produces a magnetic field, above the topological plane, equivalent to that of a mirror magnetic monopole positioned at a mirror-symmetric point with a universal value of induced magnetic charge $Q'_m = C\alpha c Q_e$ [2]. Here $\alpha \cong 1/137$ is the fine-structure constant and $c$ is the speed of light. Note that this response has been termed magnetic monopole [2] even though $Q'_m$ is not a real monopole because the induced magnetic field in Fig. 1a is pointing outwards above the plane and inwards below the plane such that $\nabla \cdot \boldsymbol{B} = 0$. Nonetheless, magnetic-monopole-like response has unique properties reflecting its topological nature that are fundamentally different from conventional magnetic response. In particular, the induced local magnetization $m_z(r)$, found from the relation $\boldsymbol{J}^T = \nabla \times m_z \hat{z}$, is proportional to the local electric potential $V$, $m_z(r) = \sigma_{yx} V(r)$, which represents a non-local relation between surface magnetization and electric field $\boldsymbol{E}(r) = -\nabla V(r)$ which has no counterpart in nontopological systems. Here $\sigma_{yx} = Ce^2/h$ is the quantum Hall (QH) conductance, $e$ is the elementary charge, and $h$ is Planck's constant. A striking manifestation of this nonlocality is that the total magnetic moment induced by the point electric charge, $M = \int_0^\infty m_z(r) d^2r = \int_0^\infty J_\varphi^T(r) r^2 dr$, as predicted from Eq. (1), becomes infinite, a behavior which is in sharp contrast to the conventional ME response. Remarkably, infinite magnetic moment also implies infinite energy $U = -\boldsymbol{M} \cdot \boldsymbol{B}$ that is required to generate the magnetic monopole response, which is cut off in reality by the system size or by disorder as we show below. Here $\boldsymbol{B} = B\hat{z}$ is the applied magnetic field.

The TME and the corresponding magnetic monopole response have so far eluded direct microscopic observation [11–14]. The technical challenges lie in the need of bringing in a controllable electric charge into close proximity to a topological surface and in measuring the resulting minute local magnetic response. We accomplished this by utilizing a nanoscale superconducting quantum interference device fabricated on the apex of a sharp pipette (SQUID-on-tip, SOT) [3]. Here, Pb SOTs have been deployed with a typical diameter $d = 60$ nm and scanned at a height of $h \cong 30$ nm above the sample surface at $T = 300$ mK. They possess magnetic field sensitivity of 30 nT/Hz$^{1/2}$ and spin sensitivity of 0.5 $\mu_B$/Hz$^{1/2}$ in a background applied magnetic fields $B \approx 1$ T (see SM3).

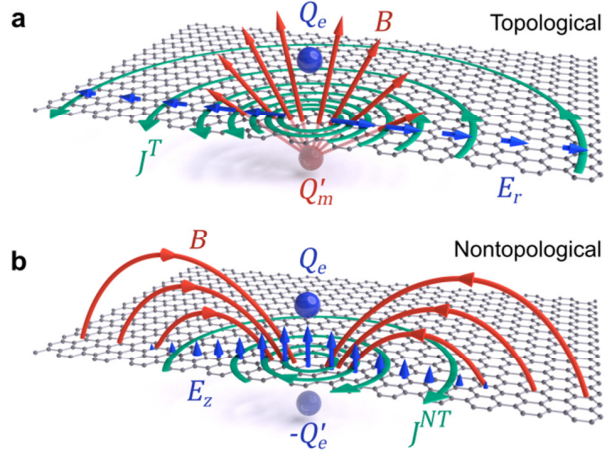

**Fig. 1. Magnetic monopole response and the topological and nontopological magnetoelectric effects.** (**a**) Schematic illustration of an electric point charge $Q_e$ (blue) creating an in-plane radial electric field $E_r$ (blue arrows) in an incompressible QH ground state of a graphene monolayer. The TME gives rise to circulating topological currents $J_\varphi^T(r) = \sigma_{yx} E_r(r)$ (green) that decay as $1/r^2$. These currents generate a magnetic field above the graphene (red arrows) of a mirror magnetic monopole with charge $Q'_m$ (transparent red). (**b**) Same configuration as (a) but in a compressible QH state in which $Q_e$ induces a mirror electric charge $-Q'_e$ (transparent blue) and an out-of-plane electric field (blue arrows). Since in this case the in-plane electric field is screened, $E_r = 0$, one expects no circulating currents. Yet, counter circulating nontopological currents $J^{NT}$ (green) arise due to nontopological ME generating an opposite magnetic field which corresponds to a magnetic dipole rather than monopole.

Our topological system of choice is hBN-encapsulated graphene in the QH state (SM1, SM2, and Figs. S1 and S2). Such van der Waals heterostructures have the advantage of exhibiting low disorder, large Landau level (LL) energy spacing, and well resolved quantum Hall states even at moderate $B$ [15,16]. This system offers exceptional density tunability and the electronic states are in close proximity to the sample surface. With the help of backgates, *p-n* junctions can be formed and the Chern numbers can be tuned in situ [17,18]. This is not possible in magnetic topological insulators that so far support only $C = \pm 1$ [19,20].

We used three different strategies to create an electric charge and the corresponding spatially dependent electric fields to explore the various regimes of the ME. The first method exploits native disorder such as charged impurities and potential variations in the sample, which induce a ME response that is spatially imaged with the SOT. This approach can provide essential information on the structure of the disorder but lacks controllability. A second method employs creation of tunable local potential in a lateral *p-n* junction induced by two separate backgates [17,18]. We used it to directly image the microscopic equilibrium currents of the QH edge states. The third method is based on the application of a non-zero voltage $V_{tg}$ to the SOT creating a controllable charge or potential at the tip apex and measuring the local ME response using the SOT.

We first discuss the TME arising due to native disorder, e.g. a point charge $Q_e$ localized at the hBN top surface (Fig. 2a). This charge, located at $(0,0,z_0)$, generates a radial electric field in the plane of graphene in the incompressible QH state, $E_r(r) = \frac{r}{4\pi\varepsilon(r^2+z_0^2)^{3/2}}Q_e$, which in turn induces a circulating current $J_\varphi^T(r) = \sigma_{yx}E_r(r)$ (SM9) and the corresponding mirror magnetic monopole $Q'_m$ according to Eq. (1) ($\varepsilon$ is the effective permittivity). The numerically calculated $J_\varphi^T(r)$ is presented in Fig. 2b vs. the global filling factor $\nu$ controlled by the backgate voltage $V_{bg}$ (SM12 and Fig. S10). As expected, $J_\varphi^T(r)$ decays with $r$ and is present only in the incompressible states with integer filling factors $\nu = C = \pm 2, \pm 6, \pm 10, \ldots$ when the Fermi level $\epsilon_F$ resides within the LL energy gaps. The magnitude of the current grows with the Chern number $|C|$ (as $\sim\sqrt{|C|}$, see SM6) and is of opposite chirality for electrons ($C > 0$) and holes ($C < 0$). In the compressible metallic regime, in contrast, $E_r(r)$ is screened and hence the topological currents are eliminated entirely. Note that these topological currents are usually referred to as chiral QH edge state currents. However, as we show below, the equilibrium QH edge states rather than carrying the commonly assumed downstream current along the chirality direction, carry pairs of counterpropagating downstream and upstream currents. These currents originate from two different mechanisms pertaining to the topological and nontopological ME and we will refer to them as $J^T$ and $J^{NT}$ respectively.

To detect the microscopic currents, we attach the SOT to a quartz tuning fork (TF) (SM4) and oscillate the SOT parallel to the graphene surface with a small amplitude $x_{ac}$ along the $\hat{x}$-direction (Fig. 2a and SM4). The SOT thereby measures the corresponding $B_z^{TF}(x,y) = x_{ac}\partial B_z(x,y)/\partial x \propto J_y(x,y)$ (Fig. 2h) and provides a direct visualization of the $y$ component of the local current $J_y(x,y)$ (SM5 and Fig. S3). Figure 2i shows profiles of $B_z^{TF}(x)$ upon sweeping $V_{bg}$ through the compressible and incompressible QH states. At integer filling factors $\nu = C$, the data provide the first direct local observation of the TME and a unique nanoscale rendering of the equilibrium topological currents. The observed behavior is in good qualitative agreement with the described model (Fig. 2b), displaying $J_y^T(x) = \sigma_{yx}E_x(x)$ of varying sign with $x$, which grows with $|C|$ and changes the overall sign upon crossing the zeroth LL.

Strikingly, in addition to the topological currents in the incompressible states, large currents are also observed in Fig. 2i in partially filled LLs, as further exemplified in the two-dimensional image in Fig. 2h. Naively, no currents are expected in this compressible, metallic regime, due to screening of the in-plane electric field. Surprisingly, these currents are of opposite polarity to those in the incompressible state (see the color inversion) and of comparable amplitude. Note that partial screening of the electric field due to finite compressibility, will result in weaker currents, and cannot explain the observed polarity inversion, since the polarity of $\sigma_{yx}$ and the direction of the external electric field will not change. We now show that these observed equilibrium currents are of an entirely different nature than the topological currents [4] and manifest a nontopological ME. These $J^{NT}$ currents apparently do not couple directly to transport measurements [21] and moreover, since they do not arise from the relation $\boldsymbol{J} = \sigma \boldsymbol{E}$, they are invisible to scanning probe techniques that do not measure currents directly, such as Kelvin probe [22], SET [23], and scanning capacitance [24]. As a result, even though their existence was predicted theoretically [4], these currents remained undetected so far. Our scanning SQUID-on-tip technique allows the first direct nanoscale imaging of these equilibrium currents.

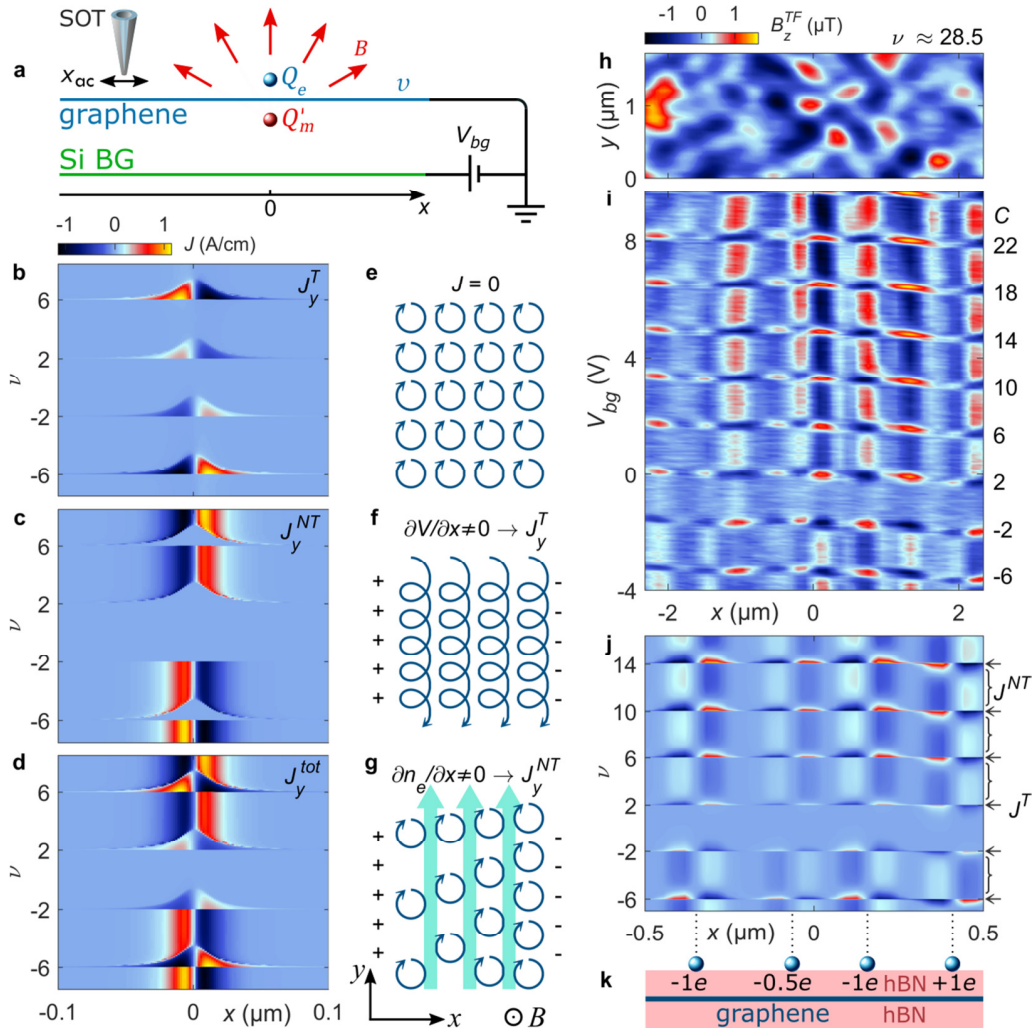

**Fig. 2. Topological and nontopological equilibrium currents in graphene in the presence of charge disorder.** (**a**) Schematic setup showing an impurity charge $Q_e$ (blue) inducing a response magnetic field $B$ (red) in a form of mirror magnetic monopole $Q'_m$ in the incompressible state. The backgate $V_{bg}$ controls the graphene global filling factor $\nu$. (**b**) Numerical simulation of $J_\varphi^T(r)$ vs. global $\nu$ for a charge $Q_e = -0.5e$ positioned on top of the hBN surface 15 nm above the graphene. The plot shows the cross section of $J_y^T(x)$ through the origin. (**c**) Same as (b) but for $J_\varphi^{NT}(r)$ that flows only in the compressible states with opposite chirality. (**d**) The total current $J_\varphi = J_\varphi^T + J_\varphi^{NT}$. (**e**) Semiclassical picture of cyclotron orbits of holes with mutually canceling neighboring currents resulting in zero total current. (**f**) In the presence of an in-plane electric field $E_x$ (+ and − signs) the cyclotron orbits acquire a drift velocity resulting in a non-zero $J_y^T$ in the incompressible state. (**g**) In the compressible regime the external in-plane electric field is screened by establishing a charge density gradient, giving rise to $J_y^{NT}$ flowing in opposite direction (cyan arrows). (**h**) Experimental $B_z^{TF}(x,y) \propto J_y(x,y)$ for $\nu \approx 28.5$ compressible state revealing equilibrium $J^{NT}$ due to native charge disorder. (h-j) share the same color bar. (**i**) Experimental $B_z^{TF}(x) \propto J_y(x)$ vs. $V_{bg}$ revealing alternating bands of $J^T$ and $J^{NT}$ due to native disorder. Right axis labels indicate the corresponding bulk $C$. (**j**) Numerically calculated $B_z^{TF}(x)$ vs. $\nu$ induced by four charges in (k). Curly brackets and arrows indicate bands of $J^{NT}$ and $J^T$ respectively. (**k**) Illustration of four impurity charges on hBN surface used for calculation in (j). See SM14 for parameters.

The following semiclassical picture is instructive to distinguish the origin of the nontopological and topological currents (SM6). In strong magnetic fields and in the absence of an in-plane electric field, the electron or hole cyclotron orbits can be described semiclassically as an array of circles resulting in zero average current (Fig. 2e). In the incompressible state, an in-plane electric field along the x- direction, $E_x = -\partial V/\partial x$, causes the orbitals to convert into spirals drifting along the $y$-direction and generating the current $J_y^T = \sigma_{yx} E_x$, as shown schematically in Fig. 2f. The topological nature of these equilibrium currents manifests itself in the fact that $\sigma_{yx}$ is quantized. On the other hand, in the compressible regime carriers redistribute themselves and screen the in-plane electric field. As a result $J_y^T$ vanishes, but at a cost of a non-zero gradient in carrier density, $\partial n_e/\partial x$ (Fig. 2g). Since each orbital carries a magnetic moment [25] $\boldsymbol{\mu}_e = -v_F\sqrt{|e\hbar n/2B|}\hat{z}$ that results in local magnetization $\boldsymbol{m} = |n_e|\boldsymbol{\mu}_e$ ($v_F$ being the Fermi velocity and $n$ the LL index), the induced $\partial n_e/\partial x$ causes gradients in $\boldsymbol{m}$, and hence produce equilibrium currents through [4] $\boldsymbol{J}^{NT} = \nabla \times \boldsymbol{m}$. This accounts for a non-zero $J_y^{NT} = \mu_e \partial |n_e|/\partial x$ in Fig. 2g which flows in the direction opposite to the topological current $J_y^T$ in Fig. 2f. Alternatively, $J_y^{NT}$ can be understood as arising from uncompensated contributions to the current from neighboring orbitals in the presence of a gradient in the orbital density (Fig. 2g). The $J^{NT}$ resemble magnetization currents in magnetic materials. Note, however, that here they do not arise as a diamagnetic or paramagnetic response to an applied magnetic field but rather as a nontopological magnetoelectric response to an applied electric field and hence can be of either polarity and of controllable magnitude. The corresponding quantum mechanical origin of both $J^T$ and $J^{NT}$ is described in SM6, with $J^{NT}$ arising from the non-homogenous distribution of the expectation value of the quantum mechanical current operator $\langle \hat{J}(x) \rangle$ (Fig. S4). Figure 2c depicts $J^{NT}$ generated by a point charge configuration of Fig. 2a, while the total current, $J^{tot} = J^T + J^{NT}$, is plotted in Fig. 2d. The calculated $B_z^{TF}(x)$ along a line crossing four point charges (Figs. 2k and S10) is presented in Fig. 2j and qualitatively reproduces the alternating $J^T$ and $J^{NT}$ stripes observed experimentally in Fig. 2i.

The topological and nontopological magnetoelectric effects in the QH regime display a remarkable duality (Fig. 1). An external charge $Q_e$ induces a mirror electric charge $-Q_e'$ in a compressible state (Fig. 1b) and a mirror magnetic charge $Q_m' = C\alpha c Q_e$ in the incompressible phase (Fig. 1a). Both types of mirror charges give rise to circulating currents: $J^{NT}(r)$ originates from an out-of-plane component of the electric field $E_z(r)$ that determines $n_e(r)$ and $\nabla n_e(r)$, while $J^T$ stems from the in-plane component $E_r(r)$. However, the two types of current are fundamentally different. In the nontopological case the magnetoelectric response to the charge $Q_e$ induces currents that decay as $J^{NT}(r) \propto r^{-4}$ generating a total magnetic moment $M^{NT} = \pi \int_0^\infty J^{NT}(r) r^2 dr = -Q_e \mu_e/e$, which is finite and local in its nature. In contrast, the topological response creates $J^T(r) \propto r^{-2}$, resulting in infinite magnetization $M^T$ for any electric charge $Q_e$. This nonlocal diverging response is the hallmark of a topological phase with a broken time reversal symmetry and accounts for the appearance of a mirror magnetic monopole rather than a magnetic dipole. Depending on the parameters, $J^T$ and $J^{NT}$ may become comparable (SM6 and Fig. S5), but they are always of opposite chirality as shown in the numerical simulations (Figs. 2b-d and 2j) and clearly observed in the experimental data in Fig. 2i. As $V_{bg}$ is varied, the measured patterns in the adjacent compressible and incompressible bands have comparable magnitude but are of opposite sign, and the overall magnitude of $J^T$ and $J^{NT}$ grows with $n$ (SM6). Remarkably, the zeroth LL shows essentially no nontopological $J^{NT}$ in Fig. 2i, as indeed predicted theoretically (SM6 and Figs. 2c and 2j).

Rather than relying on the static native disorder, we can create a variable potential by adding a graphite backgate $V_{bg}^L$ to the left half of the sample (Fig. 3a). By varying $V_{bg}^L$ and the Si backgate $V_{bg}^R$, we can create a *p-n* junction with a tunable in-plane electric field $E_x$. Due to the charge density gradient, alternating compressible and incompressible stripes should thus be formed across the junction [4,22,26]. Utilizing this setting, we present here the first direct nanoscale imaging of the equilibrium currents in the QH edge states. Figures 3b-e show a sequence of images of $B_z^{TF}(x,y) \propto J_y(x,y)$ as the filling factors $\nu_L$ and $\nu_R$ are increased with opposite polarities by sweeping $V_{bg}^L$ and $V_{bg}^R$ (see SM7 and movie M1 for the full sequence). For $|\nu_L|, |\nu_R| < 2$ both sides of the junction are in the $n = 0$ LL (Fig. 3b) and no edge currents are present (in contrast to higher LLs, see Fig. S6). Once $n = \pm 1$ LLs are reached, an incompressible strip carrying topological current $I^T$ is formed on each side. The two incompressible strips carry $I^T$ in opposite directions following the downstream chirality of the QH edge states as expected (blue arrows in Fig. 3c). As the potential difference across the junction is increased they move towards the center. The remarkable observation, however, is that as $\nu_L$ and $\nu_R$ are further increased, counter propagating currents adjacent to $I^T$ are formed (red arrows in Fig. 3d). These currents, which flow upstream against the QH chirality, are the result of the nontopological magnetoelectric contribution and are commonly ignored. This pair of counter propagating equilibrium currents, $I^T$ and $I^{NT}$, nearly cancel each other, so that the total edge current does not grow with further addition of LLs but merely oscillates (Fig. 3e). This mixed topological and nontopological magnetoelectric behavior is more clearly resolved by inspecting line cuts across the junction (Fig. 3f) forming a tree-like pattern of $I^T$ and $I^{NT}$ pairs moving towards the center with increasing filling factor difference across the junction, in agreement with numerical simulations in Figs. 3g-j. Previous scanning probe studies could differentiate between compressible and incompressible regions [22,24,27,28] but the presence of $I^{NT}$ was never uncovered.

By depleting carriers in the left side of the sample ($\nu_L = 0$) we can form fully controllable electrostatically defined QH edge states. This allows us to directly image the evolution of $I^T$ and $I^{NT}$ within individual QH edge states by varying $V_{bg}^R$ as presented in SM8, Fig. S7, and movie M2. Edge states in the context of transport studies are usually considered to be formed when a fully occupied LL with a gap above it crosses the Fermi level. This happens upon approaching the sample edge where the carrier density is gradually depleted to zero (Fig. S7g). Each edge state consists of an incompressible region with varying electric potential carrying downstream $I^T$ followed by a compressible strip with constant electric potential that carries an upstream $I^{NT}$. Thus, in contrast to the common notion, each QH edge state under equilibrium conditions carries a pair of counterpropagating currents of comparable magnitude (SM6 and Fig. S7g). In presence of an external bias the global out-of-equilibrium transport current is carried by the overall net $I^T$ while the net $I^{NT}$ should remain zero (SM8). Locally, however, we expect the external bias to change the magnitude and the distribution of both $J^T$ and $J^{NT}$ within the individual QH edge states. This nonequilibrium change in the balance between the two counterpropagating currents provides new insight into the microscopic structure and dynamics of the QH edge states and may provide new means (see SM8) for local microscopic visualization of out-of-equilibrium processes of edge reconstruction and energy equilibration in integer and fractional QH states.

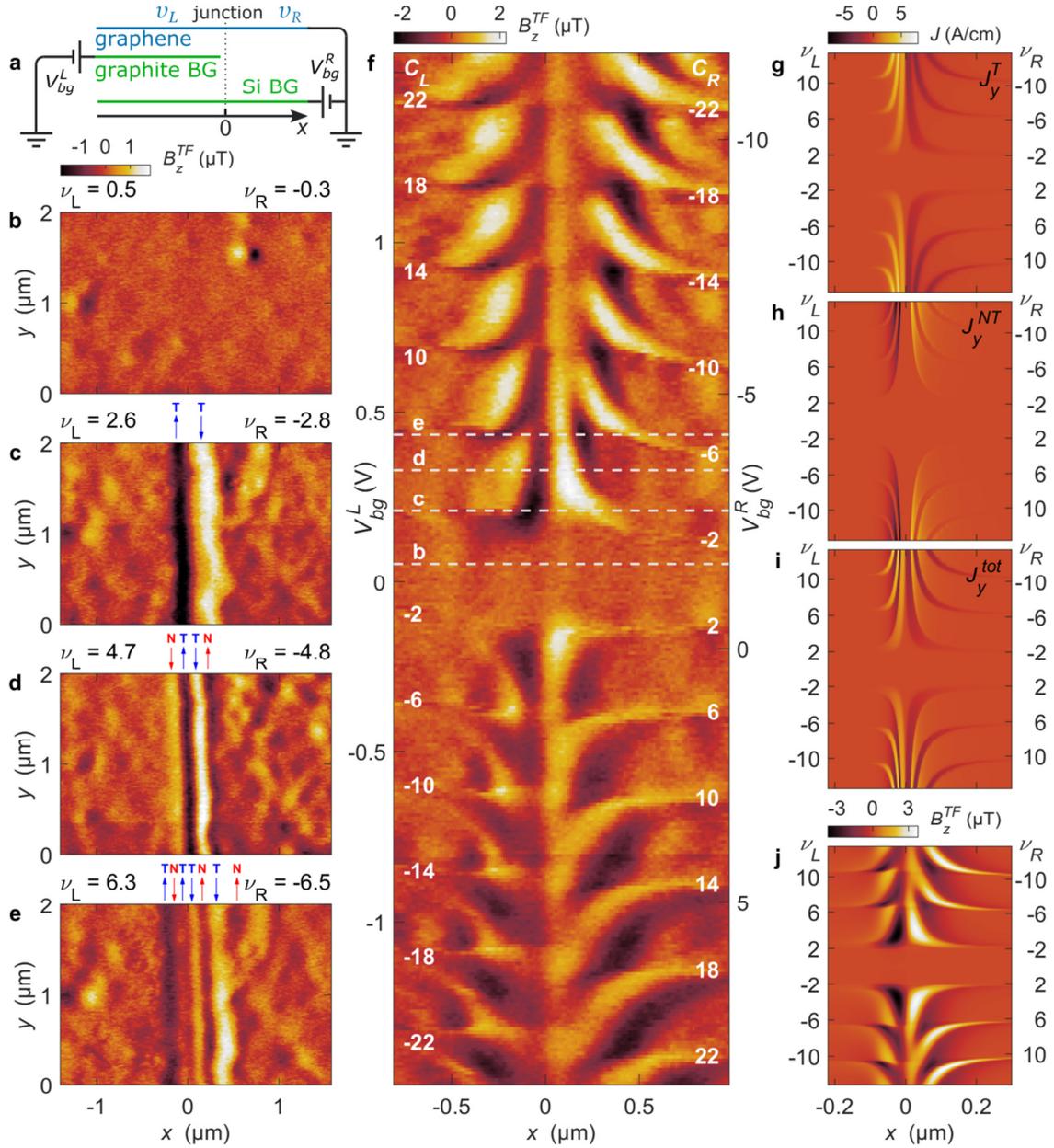

**Fig. 3. Topological and nontopological QH edge state currents in a *p-n* junction**. (a) Measurement setup with graphite and Si backgates allowing independent tuning of the left and right filling factors $\nu_L$ and $\nu_R$. (b-e) 2D imaging of $B_z^{TF} \propto J_y$ of a graphene *p-n* junction at different filling factors $\nu_L$ and $\nu_R$. Blue and red arrows mark $I^T$ and $I^{NT}$ respectively and their direction. (b) In the zeroth LL no currents are present along the junction. (c) For $\nu_L = 2.6$ and $\nu_R = -2.8$ counter propagating $J^T$ current strips appear on the *p* and *n* sides of the junction (blue arrows). (d) Upon increasing the density difference across the junction, the two $J^T$ strips get closer and subsequent strips with counter propagating current $J^{NT}$ emerge (red arrows). In (e) two additional $J_T$ strips appear. (f) Line scans across the junction sweeping $V_{bg}^L$ and $V_{bg}^R$ in opposite directions revealing the spatial evolution of $J^T$ and $J^{NT}$. Dashed lines mark the gate voltages at which the images (b-e) were recorded. (g-j) Numerical simulations reproducing the measurement in (f). They demonstrate the evolution of $J^T$ (g), $J^{NT}$ (h), total current (i), and the calculated $B_z^{TF}$ (j). See SM14 for parameters.

Employing the gained insight into the structure of the edge states we now address the general case of the monopole-like response induced by a variable charge or potential $V_{tg}$ applied to the SOT (Fig. 4a). In order to generate a mirror magnetic monopole response, the entire sample has to be in an incompressible state. Hence, the potential drop along the graphene between the SOT location and far away, $\Delta V = V_0 - V_\infty = Q_e/4\pi\varepsilon z_0$, where $z_0$ is the height of the tip above the graphene surface, should not exceed the energy gap between the LLs, $e\Delta V_{max} = \Delta E_n = E_{n+1} - E_n$, where $E_n = \text{sign}(n)\sqrt{2e\hbar v_F^2|Bn|}$. This requirement imposes an upper bound on the allowable $Q_e$ on the tip, equal to $Q_{max}^T = 4\pi\varepsilon z_0 \Delta E_n/e$. Equivalently, for the nontopological magnetoelectric effect, the maximum induced charge density variation should not surpass the density of states of a single LL, resulting in $Q_{max}^{NT} = 8\pi Be^2 z_0^2/h$. Remarkably, for our experimental configuration $Q_{max}^T$ and $Q_{max}^{NT}$ are both of the order of just a single electron charge $e$ (SM10 and Fig. S8). When $Q_e$ exceeds this value, the linear topological and nontopological magnetoelectric effects break down, giving rise to a regime where both effects get mixed. This mixed magnetoelectric effect (MME) exhibits a nonlinear response and a rich phase diagram beyond the mirror magnetic monopole framework [2,12], which has not been explored so far.

Figure 4b presents the numerically calculated $B_z$ signal at the SOT position vs. $V_{tg}$ and $V_{bg}$ revealing a complex mixed magnetoelectric phase diagram in the form of tiled diamonds. The widely searched for topological mirror magnetic monopole state [2,11–14] is present only at singular points at the vertices of the diamonds in Fig. 4b in the limit of $V_{tg} \to 0$ (see SM11 and Fig. S9). The remainder of the phase diagram is dominated by the MME phase in which each diamond corresponds to a different state classified by two quantum numbers $(n_1, n_2)$, describing the LL numbers at infinity and below the tip, respectively (Fig. 4c). The middle row of diamonds with $n_1 = n_2$ describes the nontopological magnetoelectric phase in which the entire sample is compressible. In the neighboring diamonds with $n_2 = n_1 \pm 1$ the mixed magnetoelectric effect produces a response in the form of a magnetic dipole made up of a narrow ring of current $I^T$ enclosed by counter propagating current rings of $I^{NT}$ as illustrated in Figs. 4d-h. The following rows of diamonds originate from an increasing number of concentric rings of $I^T$ and $I^{NT}$ (Fig. S9). Of particular interest are diamonds with $n_2 = -n_1$ corresponding to circular *p-n* junctions, where electron (hole) islands are formed in a hole (electron) doped sample. In this case concentric rings of $I^T$ of opposite chirality appear constituting magnetic quadrupole-like and higher order moments (Fig. S9). The quantum mechanical calculation of the MME phase diagram (SM13) is presented in Fig. S11 showing a good general agreement with the semiclassical results. Figure 4h shows a comparison of $J^{tot}$ calculated by both methods. The quantum mechanical QH wavefunction extending over a few magnetic lengths $l_B = \sqrt{\hbar/eB} \approx 25$ nm causes smoothing of the sharp current variations seen in the semiclassical calculation emphasizing the common quantum mechanical origin of $J^T$ and $J^{NT}$ (Fig. S12).

In order to increase the signal to noise ratio, rather than measuring $B_z$, we measure $B_z^{ac} = V_{bg}^{ac} \partial B_z / \partial V_{bg}$ by adding a small *ac* modulation $V_{bg}^{ac}$ to $V_{bg}$, as illustrated in Fig. 4a (SM5). The experimental data in Fig. 4i are in good agreement with the calculated $B_z^{ac}$ in Fig. 4c revealing the complex phase diagram of the MME (SM11). In addition, some fine structure in $B_z^{ac}$ is resolved within the diamonds in Fig. 4i which can be ascribed to the bound electronic states within the tip-induced "wedding cake" potential (SM13) recently observed in STM studies [29–31].

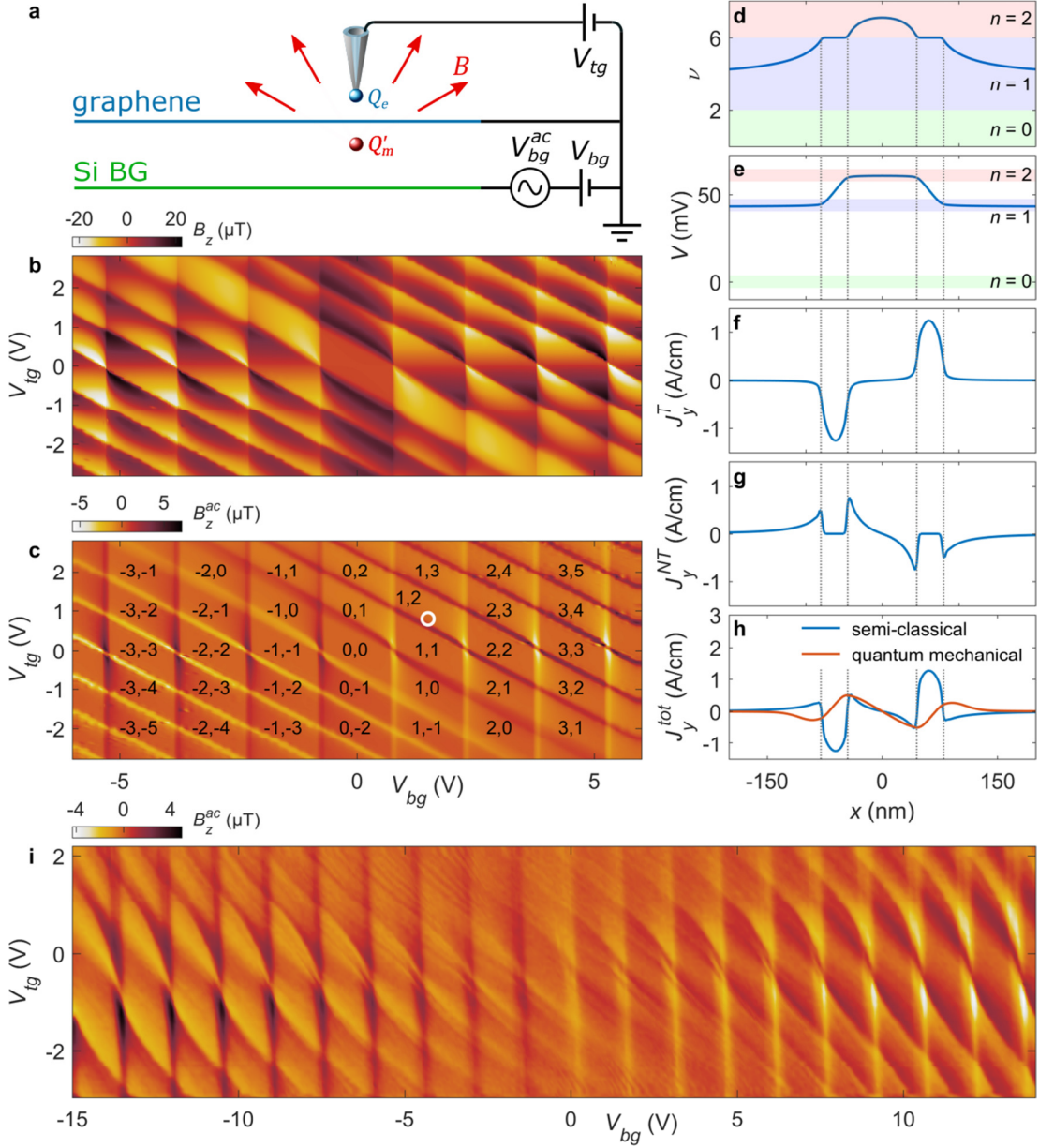

**Fig. 4. Mixed magnetoelectric effect**. (**a**) Schematic experimental setup showing charge $Q_e$ on the SOT inducing a response of mirror magnetic monopole $Q'_m$. (**b**) Numerical simulation of $B_z$ at the location of the SOT vs. $V_{bg}$ and $V_{tg}$. (**c**) Calculated $B_z^{ac}$ in response to $V_{bg}^{ac}$ reproducing the experimental results in (i). Diamonds are characterized by quantum numbers $(n_1, n_2)$ indicating the LL far from the SOT and below it, respectively. Induced magnetic monopole response is present only at the vertices of the diamonds at $V_{tg} \cong 0$. (**d-h**) Calculated filling factor $\nu$ (**d**), potential $V$ (**e**), $J^T$ in the incompressible regions (**f**), $J^{NT}$ in the compressible regions (**g**), and the total current $J^{tot} = J^T + J^{NT}$ (**h**) vs. position $x$ with the SOT stationed at the origin, for $V_{bg} = 1.5$ V and $V_{tg} = 0.8$ V (white circle in (c)). (**i**) Color rendering of the experimentally measured $B_z^{ac}$ in response to the $ac$ excitation $V_{bg}^{ac}$ vs. $V_{bg}$ and $V_{tg}$ revealing the rich phase diagram of the MME. See SM14 for parameters.

Imaging equilibrium currents at the nanoscale has the potential to be used as a novel probe of the internal structure of the quantum mechanical currents and wavefunctions. The revealed intricate configurations of topological and nontopological magnetoelectric currents in the bulk and at the edges shed light on the microscopic mechanisms of charge and heat flow in the quantum Hall edge states and on the coexistence and stability of topological phases in the presence of disorder. In addition to providing the first direct imaging of the equilibrium-state currents, the developed technique can be readily expanded to imaging out-of-equilibrium currents for revealing edge transport mechanisms in quantum spin Hall, quantum anomalous Hall, and equilibration processes in complex edge reconstruction structures in integer and fractional QH states.

**Acknowledgments**

We thank Leonid S. Levitov and Andrey V. Shytov for stimulating discussions and M. E. Huber for the SOT readout setup. This work was supported by the European Research Council (ERC) under the EU Horizon 2020 program grant No 785971, by the Israel Science Foundation grant No. 921/18, by NSF/DMR-BSF Binational Science Foundation (BSF) grant no. 2015653, and by the Leona M. and Harry B. Helmsley Charitable Trust grant 2018PG-ISL006. J.H.S. is grateful for financial support from the graphene flagship. C.K.L. acknowledges support from the STC Center for Integrated Quantum Materials (CIQM) under NSF award 1231319. C.K.L. and E.Z. acknowledge the support of the MISTI (MIT International Science and Technology Initiatives) MIT–Israel Seed Fund. Y.K. thanks the Humboldt Foundation. The growth of hexagonal boron nitride crystals was sponsored by the Elemental Strategy Initiative conducted by the 497 MEXT, Japan and the CREST (JPMJCR15F3), JST.

# Nanoscale imaging of equilibrium quantum Hall edge currents and of the magnetic monopole response in graphene

Aviram Uri[1*], Youngwook Kim[2], Kousik Bagani[1], Cyprian K. Lewandowski[3], Sameer Grover[1], Nadav Auerbach[1], Ella O. Lachman[1†], Yuri Myasoedov[1], Takashi Taniguchi[4], Kenji Watanabe[4], Jurgen Smet[2], and Eli Zeldov[1*]

[1]*Department of Condensed Matter Physics, Weizmann Institute of Science, Rehovot 7610001, Israel*

[2]*Max Planck Institute for Solid State Research, D-70569 Stuttgart, Germany*

[3]*Department of Physics, Massachusetts Institute of Technology, Cambridge, MA 02139, USA*

[4]*National Institute for Material Science, 1-1 Namiki, Tsukuba, 305-0044, Japan*

[†]*Current address: Department of Physics, University of California, Berkeley, CA 94720, USA*

[*]*Corresponding authors*

**SM1.   Device fabrication**

Three graphene based van der Waals heterostructures were measured (Fig. S1). All devices consisted of an hBN/graphene/hBN stack placed on top of the 300 nm thick $SiO_2$ layer of a thermally oxidized doped silicon wafer, acting as a backgate. A graphitic layer was placed under part of the stack, serving as an additional backgate. The two backgates allowed to induce an interface of two different filling factors, $\nu_L$ and $\nu_R$, at the boundary of the graphitic layer (Fig. 3a). The van der Waals stacking of device A, was carried out with the viscoelastic transfer method as explained in Ref. [32]. Device B and C were created with the ELVACITE based pick-up method reported in Refs. [32,33]. In order to minimize the SOT distance to graphene, we used a relatively thin top hBN layer with a thickness of approximately 8 nm (devices A and C) and 11.5 nm (device B). The bottom hBN layer was 23 nm (device A) and 50 nm (devices B and C). The graphite backgate layer had a thickness of approximately 5 nm. The heterostructures were annealed in an Ar/$H_2$ forming gas atmosphere at 500°C to remove bubbles and wrinkles prior to further processing. Patterning was performed using electron beam lithography and etching as described in Ref. [34]. Contacts and leads were fabricated by thermal evaporation of a 10 nm thick Cr adhesion layer followed by a 50-70 nm Au layer. The SOT scanning studies require an exceptionally clean surface. To ensure this, extra cleaning steps were carried out. After lift-off, devices were re-annealed at 350°C. Contact mode atomic force microscopy was deployed to sweep off PMMA residues [35,36], at a scanning speed between 0.4 and 0.6 Hz and with a tip force between 50 and 150 nN, depending on the device shape and the residues height.

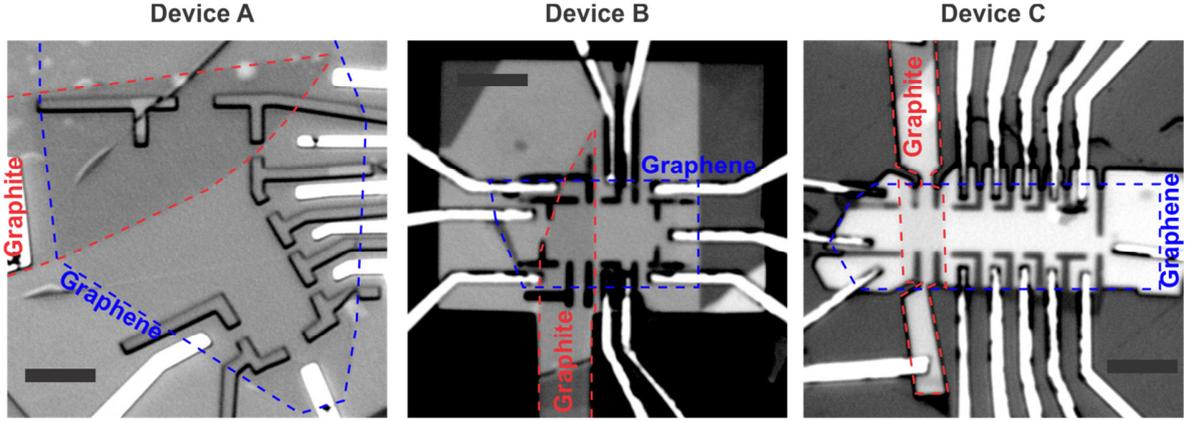

**Fig. S1. Optical microscope images of measured devices.** Red and blue dashed lines mark the graphite backgates and graphene, respectively. Etched regions appear black (devices A and B) or dark gray (device C) and the metal contacts are bright. The scale bars correspond to 5 μm.

**SM2.    Transport data**

Four-point transport measurements were performed using standard lock-in techniques with a bias current of $I = 100$ nA rms at 7 Hz. From low magnetic field data we calculated the mobility $\mu = \frac{1}{en_e\rho_{xx}}$ and mean free path $l_{mfp} = \frac{1}{2k_F\rho_{xx}}\frac{h}{e^2}$, where $n_e = \left[e\frac{d\rho_{xy}}{dB}\right]^{-1}$ is the carrier density, $\rho_{xx} = \frac{V_x}{I}\frac{W}{L}$ is the longitudinal four-point resistivity, $\rho_{xy} = \frac{V_H}{I}\frac{W}{L}$ is the Hall resistivity, $V_x$ and $V_H$ are the longitudinal and transverse voltages respectively, $L$ and $W$ are the length and width of the sample between the relevant transport contacts respectively, and $k_F = \sqrt{\pi n_e}$ is the Fermi wave-vector. Our devices showed a mean free path $l_{mfp}$ of 1 to 8 μm and a mobility $\mu \approx 10^5 - 10^6$ cm²/Vs in the Si gated regions. The graphite gated regions displayed lower mobility $\mu = 10^4 - 10^5$ cm²/Vs and $l_{mfp}$ of 0.05 to 1 μm. Figure S2 shows a Landau fan diagram of device C in the Si gated region.

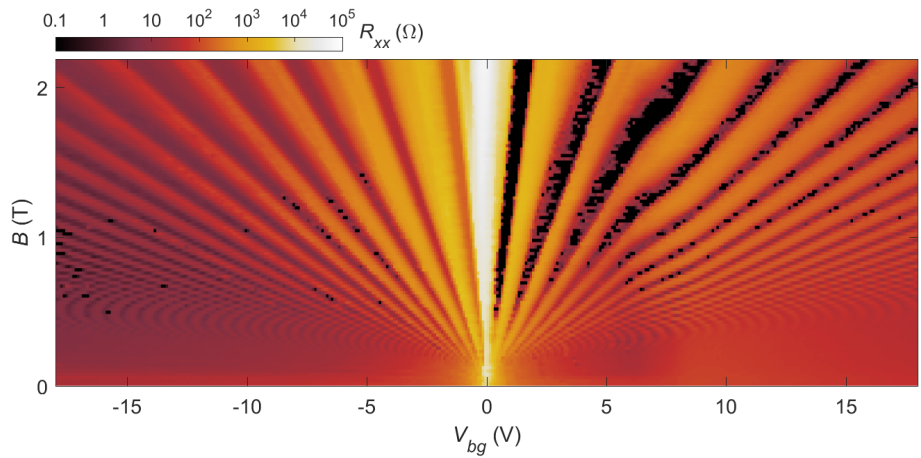

**Fig. S2. Transport measurements.** Color rendering of $R_{xx}$ of the Si gated region in device C as a function of the backgate voltage $V_{bg}$ and the applied perpendicular magnetic field $B$.

### SM3. SOT fabrication and characterization

The Pb SOTs were fabricated as described in Ref. [3] with diameters ranging from 50 to 80 nm and included an integrated shunt resistor on the tip [37,38]. The magnetic imaging was performed in a $^3$He system [39,40] at 300 mK at which the Pb SOTs can operate in magnetic fields up to 1.8 T, directed along $\hat{z}$, perpendicular to the SQUID loop. At fields $B \approx 1$ T used in this study, the SOTs displayed flux noise down to 50 n$\Phi_0$/Hz$^{1/2}$, spin noise of 0.5 $\mu_B$/Hz$^{1/2}$, and field noise down to 30 nT/Hz$^{1/2}$. Here, $\Phi_0 = h/2e$ is the two-electron flux quantum.

### SM4. Tuning fork

For height control we attached the SOT to a quartz tuning fork (TF) as described in Ref. [37]. The tuning fork was electrically excited at the resonance frequency of ~33 kHz. The current through it was amplified using a home-built trans-impedance amplifier, designed based on Ref. [41] and measured using a lock-in amplifier. The scanning was performed at a constant height of 20 to 50 nm above the surface of the top hBN. The tuning fork was vibrated along the $\hat{x}$ direction, causing the SOT to vibrate with it with a controllable amplitude $x_{ac}$ in the range of 20 to 100 nm rms. In addition to the height control, we exploited the SOT vibration to acquire the spatial derivative of the local $B_z$ field, $B_z^{TF} = x_{ac}\partial B_z/\partial x$ using a lock-in amplifier as described below.

### SM5. Modulation techniques

In order to avoid the $1/f$ noise of the SOT that is present at frequencies below ~1 kHz, we acquired *ac* signals due to two types of modulation instead of measuring the local *dc* $B_z(x,y)$.

*Backgate modulation $B_z^{ac}$*

We applied a small *ac* excitation to the backgate (Fig. 4a), $V_{bg} = V_{bg}^{dc} + V_{bg}^{ac}\cos(2\pi ft)$, where $f \cong 5$ kHz. The corresponding $B_z^{ac} = V_{bg}^{ac}\partial B_z/\partial V_{bg}$ was then measured by the SOT using a lock-in amplifier.

*Spatial modulation $B_z^{TF}$ allowing direct current imaging*

The advantage of the tuning fork induced spatial modulation of the SOT position $x_{ac}$ is that it provides a convenient means for direct imaging of the $\hat{y}$ component of the local current density $J_y(x)$. Consider a current element $J_y$ flowing in a long and narrow strip of width $\Delta x$ carrying a total current $I_y = \Delta x J_y$ in the $-\hat{y}$ direction (Fig. S3a). The magnetic field $B_z(x)$ generated by the current and measured at height $h$ above the current plane is described by the Biot Savart law. For heights $h > \Delta x$ the $B_z(x)$ is essentially governed only by the total current $I_y$ in the strip, independent of $\Delta x$ (Fig. S3b). The $B_z(x)$ is an antisymmetric function with a steep slope above the current strip. The spatial derivative $\partial B_z/\partial x$ has a sharp peak at the strip location with a height proportional to $I_y$ and a width determined by the scanning height $h$ (Fig. S3c). Measuring $B_z^{TF} = x_{ac}\partial B_z/\partial x$ thus provides a convenient method for direct visualization of the spatial current distribution $J_y(x)$ with a resolution limited by the scanning height and the SOT diameter as demonstrated by the simulation of three counterpropagating current strips in Figs. S3d-f.

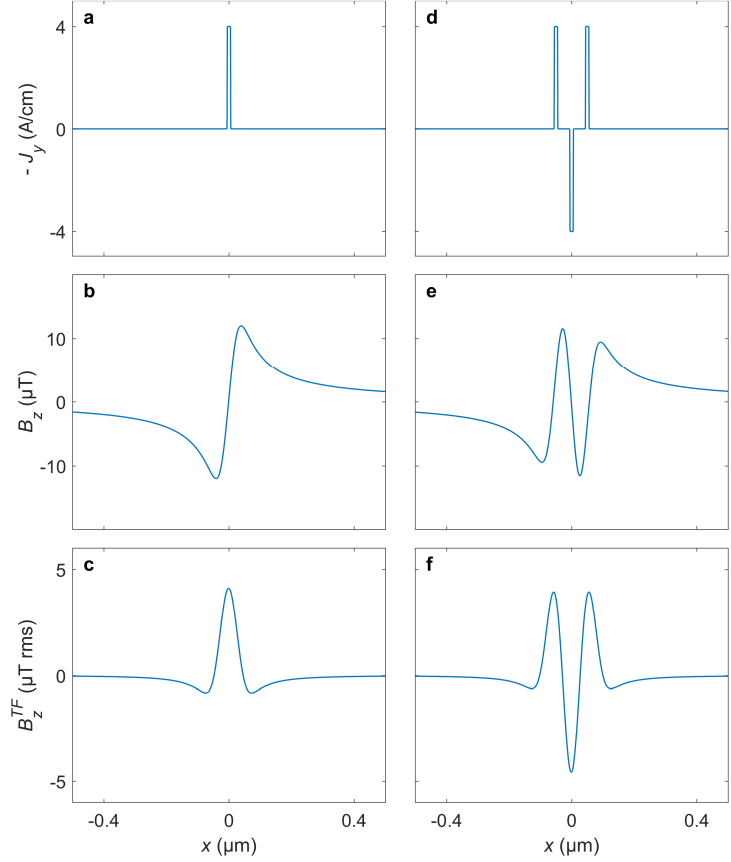

**Fig. S3. Numerical simulations demonstrating direct current imaging of $J_y(x)$ by measuring $B_z^{TF}(x)$. (a)** Current distribution $J_y(x)$ of a 10 nm wide channel carrying $I_y = 4$ µA ($\cong I_{max}^T$) in the $-\hat{y}$ direction. **(b)** Calculated $B_z(x)$ at a height of 30 nm above the sample convoluted with a 60 nm diameter SOT sensing area. **(c)** Calculated $B_z^{TF}(x)$ for $x_{ac} = 20$ nm rms. **(d-f)** Same as (a-c) but for three counterpropagating currents spaced 50 nm apart.

### SM6. Origin of and relation between topological and nontopological currents

To describe the nontopological current $J^{NT}$, we first use a semiclassical description of massless Dirac fermions in graphene with Fermi velocity $v_F$, momentum $\hbar k$, and energy $\epsilon = v_F \hbar k$. In the presence of a perpendicular applied magnetic field $B\hat{z}$, an electron moves in a cyclotron orbit with a radius $R = |\hbar k/eB|$ at cyclotron frequency $\omega_c = B|v_F^2 e/\epsilon| = B|v_F e/\hbar k|$ [25]. This cyclotron motion gives rise to a magnetic moment that we denote as $\boldsymbol{\mu_e} = \mu_e \hat{z}$. It is determined by the current $I = -e\omega_c/2\pi$ times the cyclotron area $A = \pi R^2$,

$$\mu_e = -e\frac{\omega_c R^2}{2} = -\frac{|v_F \hbar k|}{2B} = -\frac{|\epsilon|}{2B}.$$

This magnetic moment of a Dirac particle is half of the moment of a massive 2D particle, $\mu_e = -|\epsilon|/B$. The magnetic moment of an electron residing in LL $n$ at energy $\epsilon_n = \text{sign}(n)v_F\sqrt{2e\hbar|Bn|}$ is thus

$$\mu_e = -\text{sign}(B)v_F\sqrt{\left|\frac{e\hbar n}{2B}\right|}.$$

Note that electrons and holes in magnetic field have the same semiclassical diamagnetic moment and the same orbital current (opposite charge with opposite circulation direction). It is interesting to evaluate $\mu_e$ in units of Bohr magneton $\mu_B = e\hbar/2m_e$, where $m_e$ is the free electron mass:

$$\frac{\mu_e}{\mu_B} = -\text{sign}(B) m_e v_F \sqrt{\left|\frac{2n}{e\hbar B}\right|}.$$

With $v_F = 10^6$ m/s, we obtain $\mu_e \approx -313.4 \mu_B \text{sign}(B)\sqrt{|n/B\,[\text{T}]|}$, where $B$ is expressed in units of T.

A uniform number density of electrons $n_e$ will result in a uniform magnetization $\boldsymbol{m} = |n_e|\mu_e \hat{z}$ which creates no net local current and, hence, will locally be invisible to our SOT. In contrast, an inhomogeneous carrier density will generate local currents due to gradients in $n_e(x,y)$. In particular, in the compressible state, an external in-plane electric field will be screened by charge redistribution giving rise to gradients in $n_e$, and hence in current density through the relation

$$\boldsymbol{J}^{NT} = \nabla \times \boldsymbol{m} = \mu_e \left(\frac{\partial |n_e|}{\partial y}\hat{x} - \frac{\partial |n_e|}{\partial x}\hat{y}\right),$$

where the currents in $\hat{y}$ are produced by density gradients in $\hat{x}$ and vice versa. In case of cylindrical symmetry, the circulating current is given by $\boldsymbol{J}^{NT} = J_\varphi^{NT}\hat{\varphi} = -\mu_e \frac{\partial |n_e(r)|}{\partial r}\hat{\varphi}$.

Consider an interface between two regions with different densities $n_e^L$ at $x < 0$ and $n_e^R$ at $x > 0$ within the same LL (Fig. S4a-c). Within the above semiclassical picture such interface will carry a net current in the $\hat{y}$ direction of $I_y^{NT} = \int J_y^{NT} dx = \mu_e(|n_e^L| - |n_e^R|) = \mu_e(|\nu_L| - |\nu_R|)|B|/\phi_0$, where $\phi_0 = h/e$ is the single-electron flux quantum. In the quantum mechanical description, instead of the cyclotron orbits, a single electronic state with some $k_y$ (and corresponding cyclotron guiding center $x_0$) has a probability current density expectation value $\langle \hat{J}_y(x) \rangle$ as described in Fig. S4a with a zero total current. Figure S4b shows a superposition of probability currents for many different $k_y$ states (and corresponding guiding centers) with a change in occupation density at $x = 0$. The sum of the probability currents (Fig. S4c) is zero in the regions of constant density and has a current peak at the origin where a sharp density gradient is present, like in the semiclassical case. This non-zero current is attributed to the non-homogenous distribution of the current density $\langle \hat{J}_y(x) \rangle$. Note that in the quantum mechanical description the width of this nontopological current strip is broadened on the scale of the magnetic length $l_B = \sqrt{\hbar/e|B|}$. The maximal $I^{NT}$ that can flow along the interface within a single LL is given by the maximal $|n_e^L - n_e^R|_{max} = n_{eLL}$, where $n_{eLL} = 4|B|/\phi_0$ is the density of a full LL, or $|\nu_L - \nu_R|_{max} = 4$,

$$I_{max}^{NT} = |\mu_e| n_{eLL} = \frac{v_F}{\pi}\sqrt{\frac{2e^3|Bn|}{\hbar}}.$$

For $n = 1$ and $B = 1$ T, the $I_{max}^{NT} \cong 2.8$ μA.

In the incompressible topological state, the gradient $\nabla n_e$ equals zero, but an in-plane electric field causes an asymmetry in the single electronic state (Fig. S4d) with a finite net total current. A uniform superposition of such asymmetric states (Fig. S4e) gives rise to a topological current with a constant density $J^T$ (Fig. S4f). The maximal total equilibrium topological current that can flow in such an incompressible strip, formed where $\epsilon_F$ is in the gap between LLs $n$ and $\text{sign}(n)(|n| + 1)$, is $I_{max}^T = \sigma_{yx}\Delta E_n/e$, where $\Delta E_n = E_{|n|+1} - E_{|n|}$ is the gap energy and $E_n = \text{sign}(n)\sqrt{2e\hbar v_F^2|Bn|}$, resulting in

$$I_{max}^T = \frac{v_F}{\pi}\sqrt{\frac{2e^3|B|}{\hbar}}(2|n|+1)\left(\sqrt{|n|+1} - \sqrt{|n|}\right).$$

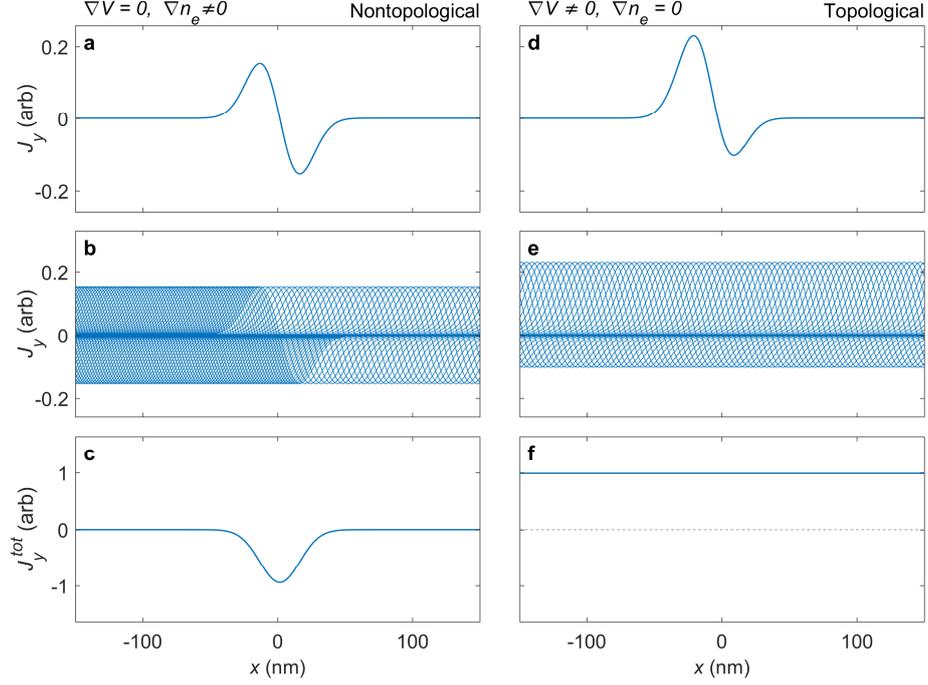

**Fig. S4. Quantum mechanical origin of $J^T$ and $J^{NT}$.** (a) Solution of the Dirac equation for the probability current density of a single $k_y$ state in the $n = 1$ LL at $B = 1.5$ T. (b) Currents of different $k_y$ states with occupation density step change from $n_e^L$ to $n_e^R$ at $x = 0$. (c) Integration of all currents in (b) giving rise to a peak in $J_y^{NT}$ around $x = 0$, due to the non-homogeneous distribution of the quantum mechanical current density of a single electronic state, with a spatial extent of the magnetic length $l_B$. (d) Current carried by a single $k_y$ state in the presence of an electric field $\boldsymbol{E} = 0.2 \times v_F B \hat{x}$ which gives rise to asymmetry and topological current. (e) Current carried by different $k_y$ states with constant occupation density $n_e$. (f) Integration of all currents in (e) showing a homogeneous non-zero $J^T$.

Here, $\sigma_{yx} = Ce^2/h$ and we took $|C| = 4|n| + 2$. The maximal topological and nontopological currents are of the same order of magnitude as shown in Fig. S5, with the exception of $n = 0$ for which $I^{NT}$ vanishes. This is because the zeroth LL has zero energy and angular momentum resulting in $\mu_e = 0$. The fact that $I_{max}^T$ and $I_{max}^{NT}$ are comparable for $n \neq 0$ may seem surprising because all filled LLs contribute to $J^T$, while $J^{NT}$ is generated only by the last occupied LL, which sustains density gradients (Fig. S7g,h). Indeed, $J^T \propto \sigma_{yx} \propto n$, however, the $I_{max}^T$ is also proportional to the energy gap $\Delta E_n$ which decreases as $\sim 1/\sqrt{|n|}$. For $J^{NT}$, the carrier density in the LLs is independent of $n$, but $J^{NT} \propto \mu_e \propto \sqrt{|n|}$. As a result, both $I_{max}^T$ and $I_{max}^{NT}$ grow as $\sim \sqrt{|n|}$.

Note that for spinless 2D particles with a parabolic dispersion, effective cyclotron mass $m$, and $E_n = \hbar\omega_c\left(|n| + \frac{1}{2}\right)$, the $I_{max}^T = \sigma_{xy}\frac{\Delta E_n}{e} = \frac{e^2|B|}{2\pi m}(|n|+1)$ and $I_{max}^{NT} = \frac{e^2|B|}{2\pi m}\left(|n| + \frac{1}{2}\right)$, with the ratio $I_{max}^{NT}/I_{max}^T = (|n| + 1/2)/(|n| + 1)$ (Fig. S5). Here, as before, we considered $I_{max}^{NT}$ for LL $n$ and $I_{max}^T$ in the incompressible state formed where $\epsilon_F$ is between LLs $n$ and $\text{sign}(n)(|n| + 1)$. The topological and nontopological currents produced by a given external perturbation are always of opposite sign as illustrated in Figs. 2f-g.

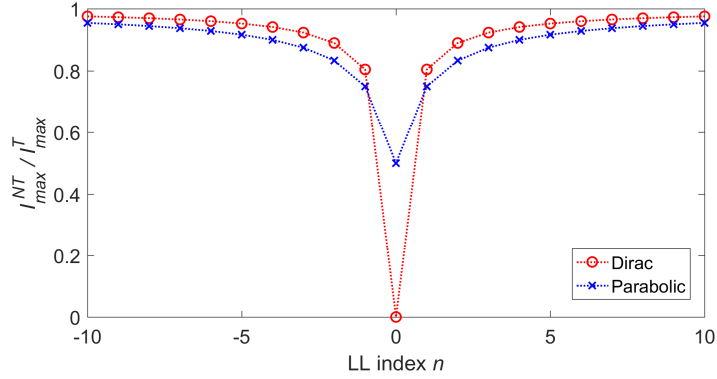

**Fig. S5. Ratio of $I_{max}^{NT}$ to $I_{max}^{T}$ vs. the LL index $n$.** Shown is the ratio of the $I_{max}^{NT}$ flowing in LL $n$ and the $I_{max}^{T}$ flowing in the incompressible strip between LL $n$ and LL $\text{sign}(n)(|n|+1)$ for the cases of Dirac dispersion in graphene (red) and parabolic dispersion relation (blue). The maximal topological and nontopological current are of the same order of magnitude, flowing in opposite directions, except for LL $n = 0$ in graphene where $I_{max}^{NT} = 0$.

To demonstrate the described equilibrium topological and nontopological currents at an interface between two regions, Fig. S6b presents line scans of $B_z^{TF}(x) \propto J_y(x)$ across the interface at fixed $y$ as the graphite and Si backgates are swept simultaneously to maintain a fixed density difference $n_e^L - n_e^R \cong n_{eLL}/2$ or $\nu_L - \nu_R = \Delta\nu \cong 2$ between the two sides (Fig. S6a). The red dashed line in Fig. S6b describes the situation where both sides of the interface are in a compressible state within the same LL $n = 6$. The common assumption is that in this case no current should be present since the entire sample is metallic with no in-plane electric fields present. Figure S6b, however, clearly shows a large peak (yellow-red) in $B_z^{TF}(x) \propto J_y(x)$ revealing the nontopological current $I_y^{NT} = \mu_e(|n_e^L| - |n_e^R|) \cong -I_{max}^{NT}/2$ flowing along the interface analogous to Figs. 2g and S4c. This current along the interface is absent in Fig. S6c for $\Delta\nu = 0$ where only the native disorder-induced currents are present. The blue dashed line in Fig. S6b shows the contrasting situation where the two sides of the interface are in different compressible states of LLs $n = 6$ and 5, separated by an incompressible strip at the interface. Within this strip, a topological current $I_y^T = I_{max}^T$ flows in opposite direction (analogous to Fig. 2f) as seen by the dark blue signal at the interface. Interestingly, since the difference in the carrier density $|n_e^L| - |n_e^R| \cong n_{eLL}/2$ is still present across the interface, in addition to the topological $I_y^T$, nontopological currents $I_y^{NT}$ (red peaks) are counterflowing on the two sides of the incompressible strip partially balancing the topological current. This feature is clearly seen in Fig. S6d, which shows the difference between the line profiles of Figs. S6b and S6c along the red and blue dashed lines. Note that the equilibrium topological current $I_y^T$ carried by an incompressible strip residing between two compressible regions (like in the case of the blue dashed line in Fig. S6b) is always equal to $I_{max}^T$ regardless of $\Delta\nu$ across the interface. In contrast, the amplitude of the nontopological current $I_y^{NT}$ (like in the case of the dashed red line) is proportional to $\Delta\nu$ and in the presented case amounts to $I_y^{NT} \cong -I_{max}^{NT}/2$. Since $I_{max}^{NT}$ and $I_{max}^T$ are of comparable magnitude (see Fig. S5), the amplitude of the nontopological current peak in Fig. S6d (red) is about half of the topological current peak (blue). By increasing $\Delta\nu$ to close to 4, comparable values of $I_y^T$ and $I_y^{NT}$ will be attained.

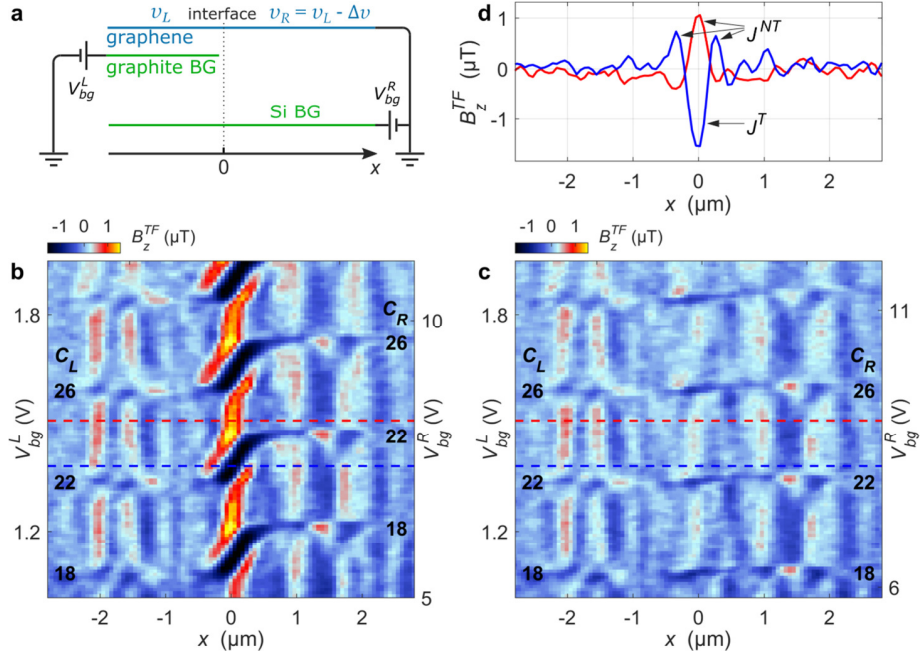

**Fig. S6**. **Demonstration of the relation between $J^T$ and $J^{NT}$.** (**a**) Experimental setup for creating an interface between two regions with carrier densities $\nu_L$ and $\nu_R$. (**b**) Line scans across the interface showing $B_z^{FT}(x) \propto J_y(x)$ as $V_{bg}^L$ and $V_{bg}^R$ are swept in the same direction maintaining a constant difference $\nu_L - \nu_R = \Delta\nu \cong 2$. Topological (dark blue) and nontopological (yellow-red) currents are periodically created at the interface. The background currents away from the interface are caused by native disorder. (**c**) Same as (b) but for $\Delta\nu \cong 0$ revealing only the disorder-induced currents. (**d**) Profiles of $B_z^{FT}(x)$ along the red and blue dashed lines in (b) after subtracting the disorder-induced profiles along the dashed lines in (c) for clarity.

**SM7.   Movie of the evolution of the magnetoelectric effect across a *p-n* junction**

The full sequence of 2d scans shown in Figs. 3b-e is given in movie M1. It shows the evolution of $J^T$ and $J^{NT}$ as $V_{bg}^L$ and $V_{bg}^R$ are swept with opposite polarities. Starting with $\nu_L \approx 0$ and $\nu_R \approx 0$ we gradually populate the left side of the graphene flake with electrons, increasing $\nu_L$ while simultaneously populating the right side with holes such that $\nu_R \approx -\nu_L$. The $\nu_L$ and $\nu_R$ indicated at the top of the frame are approximate average values calculated using the applied $V_{bg}^L$ and $V_{bg}^R$ and the estimated geometric and quantum capacitances. At low filling no currents are present since $\mu_e = 0$ in the $n = 0$ LL. When $|\nu_L|$ and $|\nu_R|$ approach 2, a strip of topological current $J^T$ appears on each side of the junction flowing in opposite directions (blue arrows) because $J_y^T = \nu \frac{e^2}{h} E_x$ and $\nu_{L,R}$ in the two incompressible strips are of opposite sign. As the filling factors increase further, counterpropagating slivers of nontopological current $J^{NT}$ appear on the outer sides of the $J^T$ strips, gradually increasing in magnitude, as $\nabla n_e$ increases. On approaching $|\nu_{L,R}| = 6$, a second pair of $J^T$ sets in and moves towards the junction. The increasing bending of the LLs makes the edge states narrower and denser. Eventually the edge states appear as if they merge when their distance falls below our spatial resolution, limited by the SOT diameter and height above graphene. Note that the potential variation is steeper on the left side because of the closer proximity of the graphite backgate to the graphene (~50 nm) as compared to the Si backgate (~300 nm, see Fig. 3a). As a result, the edge states on the left side are denser. This is clearly visible in Fig. 3f and the numerical simulations in Figs. 3g-i.

**SM8.   Electrostatically-defined quantum Hall edge states**

Using the experimental setup of Fig. 3a, setting $v_L = 0$ and varying $v_R$, we can create the situation of electrostatically defined edge states. Figure S7 and movie M2 show the evolution of $J^T$ and $J^{NT}$ edge currents versus $v_R$. Several recent studies have shown evidence of charge accumulation along the physical edges of graphene upon changing the carrier concentration with backgate [27,42–45]. This charge accumulation apparently leads to complicated configurations of QH edge states and unconventional transport behavior [27]. An important finding here is that no such charge accumulation occurs in the case of our electrostatically defined edge. This is evidenced in Fig. S7f by the monotonic upward bending of LLs visualized by the bright traces of the incompressible strips. In the case of charge accumulation at the edge, the LLs should have a nonmonotonic behavior with downward bending before the upturn towards the edge. We postulate that this absence of charge accumulation plays a central role in the improvement of the fractional QH effect in electrostatically defined devices [46].

We now focus on the microscopic details of the QH edge evolution upon varying $v_R$. Similar to the case of a *p-n* junction, a strip of $J^T$ appears when reaching $v_R = 2$ (Fig. S7c) and moves towards the gate-defined edge as $v_R$ is increased (movie M2). It is followed by a counterpropagating $J^{NT}$ that increases with $v_R$ (Fig. S7d). This sequence of advancing $J^T$ and $J^{NT}$ pairs repeats itself (Fig. S7e) upon filling of every following LL as clearly seen in Fig. S7f. Note that every pair of full-valued $J^{NT}$ and $J^T$ for $|n| \geq 1$ carries almost no net current (Fig. S5). As a result, when the edge states become too dense we can no longer resolve them within our spatial resolution.

Figure S7g shows a simplified schematic diagram of the QH edge states [26]. Each edge state carries a pair of counterpropagating currents [4]: $J^{NT}$ flowing upstream in the outer compressible strip of the edge state and $J^T$ flowing downstream with the edge chirality along the inner incompressible region. Let us focus on the $n = 2$ QH edge state. Figure S7g depicts the situation in which the Fermi level $\varepsilon_F$ is just below $n = 3$ LL. In this case the integrated currents are maximal, $I^{NT}_{max}$ and $I^T_{max}$, and since their magnitudes are comparable (Fig. S5) the overall current flowing in the edge state is close to zero. The $I^{NT}$ arises from the fact that in the incompressible strip the carrier density in the $n = 2$ LL decreases from a fully occupied to an empty level. The $I^T$, in contrast, is caused by the potential difference across the incompressible region which drives the current in all of the three underlying occupied LLs. Upon decreasing $\varepsilon_F$ within the gap, $I^T$ decreases while $I^{NT}$ remains unchanged. Since $I^T$ is proportional to $\varepsilon_F - \varepsilon_n$, where $\varepsilon_n$ is the energy of the highest underlying occupied LL, measurement of the local $I^T$ thus provides direct information on the local value of the electrochemical potential, which becomes essential under nonequilibrium conditions as discussed below. When $\varepsilon_F$ reaches just above the $n = 2$ LL, the downstream $I^T$ in this QH edge state vanishes and only the upstream $I^{NT}$ remains (Fig. S7h). In this situation the total equilibrium current carried by the $n = 2$ QH edge state reaches its maximum value given by the upstream $I^{NT}_{max}$. Decreasing $v_R$ further will start depleting the $n = 2$ LL, correspondingly reducing $I^{NT}$ until it vanishes upon the full depletion of the $n = 2$ LL. Similarly to $I^T$, such local measurement of $I^{NT}$ thus provides key information on the occupation level of the LL in a nonequilibrium case. The above process will then repeat itself for the $n = 1$ QH edge state. Thus the net equilibrium current in the QH edge states oscillates periodically with $V_{bg}$. Note that the $n = 0$ QH edge state in graphene is an exception since it carries only $I^T$ while $I^{NT} = 0$. This absence of $I^{NT}$ in the $n = 0$ LL is clearly seen in Fig. S7f and movie M2. This is not the case for 2DEG of massive particles like in GaAs heterostructures where all the LLs carry pairs of counterpropagating $I^{NT}$ and $I^T$ (Fig. S5).

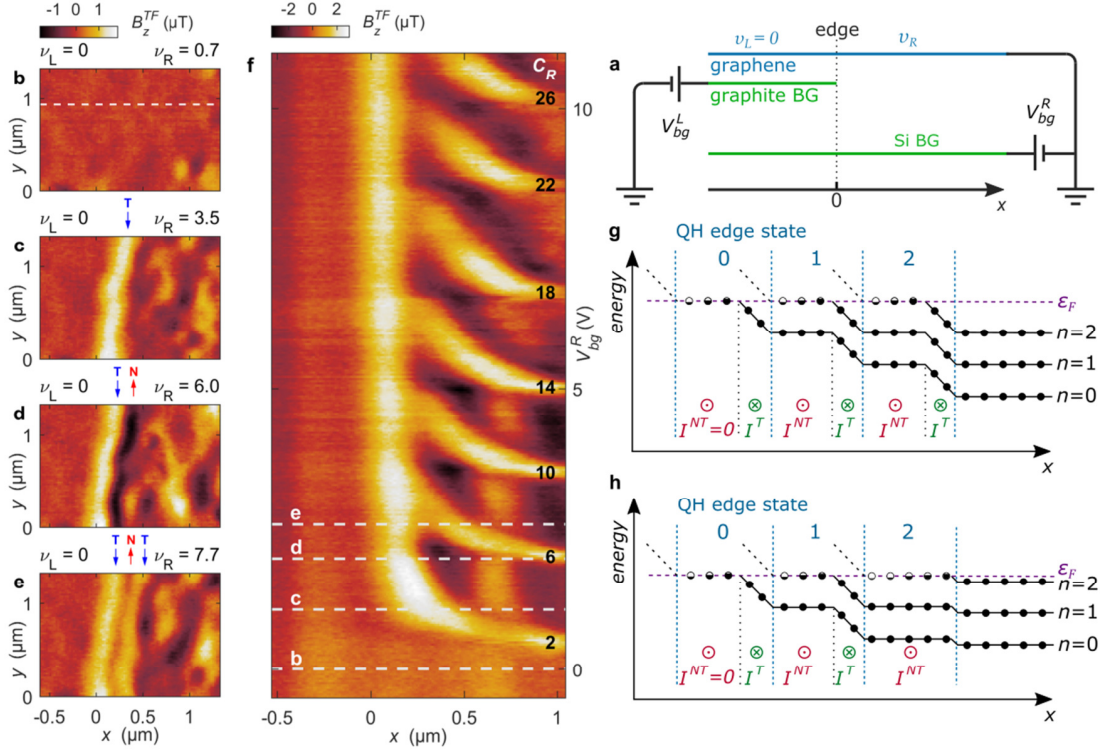

**Fig. S7**. **Imaging of quantum Hall edge currents.** (**a**) Schematic experimental setup. The graphite backgate is set to $\nu_L = 0$ to form an electrostatically defined edge and the Si backgate is used to sweep $\nu_R$. (**b**)-(**e**) $B_z^{TF}(x,y) \propto J_y(x,y)$ for different indicated $\nu_R$. The blue and red arrows point to the locations of $J^T$ and $J^{NT}$ edge currents respectively and to their directions. (**f**) Line scans of $B_z^{TF}(x) \propto J_y(x)$ vs. $V_{bg}^R$ across the gate-defined edge along the dashed line in (b) showing the evolution of $J^T$ and $J^{NT}$ edge currents with $\nu_R$. The corresponding Chern numbers $C_R$ are indicated on the right. The dashed lines show the values of $V_{bg}^R$ at which the 2d images (b-e) were acquired. (**g**) Illustration of equilibrium QH edge states in the situation where $\epsilon_F$ resides just below the $n = 3$ LL. Each edge state is constructed from an outer compressible strip carrying $I^{NT}$ upstream and an inner incompressible strip carrying $I^T$ flowing along the downstream chirality. The $n = 0$ LL is exceptional in graphene in carrying zero $I^{NT}$. (**h**) Same as (g) for a situation where $\epsilon_F$ is lowered and resides just above the $n = 2$ LL. In this case the $n = 2$ QH edge state carries mainly $I^{NT}$ with vanishing $I^T$.

The above description of counterpropagating currents that change their relative magnitude as a function of local electrochemical potential and carrier density, provides an important new framework for exploring nonequilibrium phenomena. In the equilibrium case the total current integrated over the width of the sample must be zero for $I^T$ and for $I^{NT}$ separately. In out-of-equilibrium conditions in presence of external bias, the total $I^{NT}$ should still remain zero, because it is proportional to the difference in the carrier densities at the opposite edges of the sample, but the density at sample edges has to vanish always. The total $I^T$, in contrast, becomes finite due to the induced transverse net drop in the electrochemical potential. This information provided by global transport measurements, however, does not reveal the local microscopic out-of-equilibrium processes. Under out-of-equilibrium conditions, each QH edge state generally has a different, position dependent electrochemical potential and nonequilibrium carrier distribution driven by external voltage bias or by other external perturbations including thermal gradients and heat flow. The equilibration process due to electron-electron scattering within each state and between neighbouring edge states is a subject of extensive studies using spectroscopic transport measurements [47–49]. These measurements, however, provide information only at specific lithographically predefined locations. In addition, contacts may affect the current flow making these measurements invasive. Since the local electrochemical potential and the carrier density determine the relative magnitude and distribution of $J^T$ and $J^{NT}$ within each individual edge state, our technique may thus provide a unique imaging tool to visualize and explore nonequilibrium phenomena and the microscopic mechanisms of equilibration, edge reconstruction, and charge and heat transfer between individual QH edge states under various out-of-equilibrium conditions. This will be the subject of future work.

**SM9.   TME response to a point charge**

We show that a point electric charge $Q_e$ placed at height $z_0$ above (or below) a topological surface characterized by a quantized Hall conductance $\sigma_{yx}$, induces a magnetic field above the surface identical to that of a magnetic monopole of charge $Q'_m$ placed at height $z_0$ below the surface. In the incompressible QH state the electric field $\boldsymbol{E}$ produced by the charge $Q_e$ is unscreened and its radial component $E_r$ along the topological surface is given in cylindrical coordinates by

$$E_r(r) = \frac{Q_e}{4\pi\epsilon_0} \frac{r}{(r^2 + z_0^2)^{3/2}},$$

leading to a circulating topological current density $J^T_\varphi(r) = \sigma_{yx} E_r(r)$. The magnetic field along the $z$ axis is then obtained by integrating the Biot-Savart relation for concentric current rings with current $dI(r) = J_\varphi(r)dr = \sigma_{yx} E_r(r) dr$:

$$B_z(z) = \int_0^\infty \frac{\mu_0}{4\pi} \frac{2\pi r^2}{(r^2 + z^2)^{3/2}} \frac{\sigma_{yx} Q_e}{4\pi\epsilon_0} \frac{r}{(r^2 + z_0^2)^{3/2}} dr = \frac{\mu_0}{4\pi} \frac{Q'_m}{(|z| + |z_0^2|)^2}.$$

Here, $Q'_m = \sigma_{yx} Q_e / 2\epsilon_0$. Using $\sigma_{yx} = Ce^2/h$ we get

$$Q'_m = \frac{1}{2\epsilon_0} \frac{Ce^2}{h} Q_e = C\alpha c Q_e,$$

where, $\alpha$ is the fine structure constant, $c$ the speed of light, and $C$ the Chern number. Note that $Q'_m$ flips sign with $B_z$. In the calculations above, we took $B_z$ in the positive $\hat{z}$ direction.

**SM10.   Monopole fragility**

In order to induce a magnetic monopole the entire sample has to be in the incompressible state. Hence, the potential variation across the sample cannot exceed the energy gap $\Delta E_n/e$. Due to its low density of states and the linear dispersion, graphene possesses a much larger QH energy gap than any other known time reversal symmetry broken topological system. For an applied magnetic field $B = 1$ T, $\Delta E_1 = 36.3$ meV $\cong 420$ K. But even for such a large gap, a single electron charge trapped on the surface of hBN or located on the scanning tip is sufficient to destroy the monopole state as demonstrated numerically in Fig. S8. In this case, an incompressible state throughout the sample that is required for inducing a magnetic monopole cannot be formed at any value of $V_{bg}$. In magnetic topological insulators the energy gap is estimated to be of the order of 0.5 K [50–52], and hence native charge disorder as low as $10^{-3}e$ is sufficient to prevent the formation of a magnetic monopole. Since the magnetically doped topological insulators are found to be highly disordered [40,53], we conclude that the proposed observation of the induced magnetic monopole in them [2] is unfeasible at present.

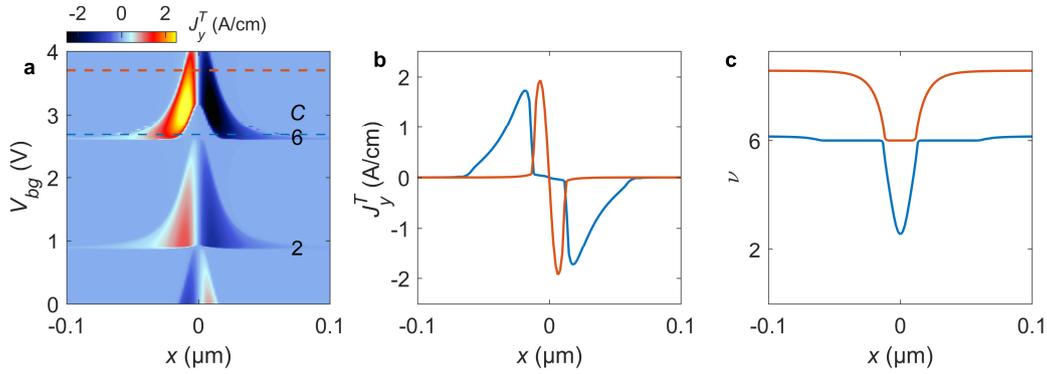

**Fig. S8. Monopole fragility.** (**a**) Numerical simulation of $J_\varphi^T(r)$ vs. $V_{bg}$ induced by a point charge of $-1.5e$ positioned on the surface of 15 nm thick hBN on graphene at $B = 1.5$ T. Shown is a cut of the topological current $J_y^T(x)$ through the origin. At these conditions, no $V_{bg}$ value exists at which the entire sample is in the incompressible state that is required for inducing a magnetic monopole. (**b**) $J_y^T(x)$ profiles along dashed lines in (A) showing that $J_\varphi^T(r)$ has always a finite extent in space in the form of a ring (blue) or a disk (orange) giving rise to a dipole-like magnetic response. In contrast, $J_\varphi^T(r)$ has to extend to infinity for a monopole response. (**c**) Corresponding filling factor $\nu(x)$ profiles showing that the incompressible states of $\nu = 6$ are confined to finite regions in the form of a ring (blue) or a disc (orange) at the origin.

### SM11. Mixed magnetoelectric state

The rich phase space of the mixed magnetoelectric effect (MME) contains a wide variety of nonlinear responses and complex configurations of topological and nontopological current flow. Figure S9 highlights three notable cases marked by colored circles in Fig. S9a. The magnetic monopole in the $C = 2$ state (Figs. S9b1-f1 and black circle in Fig. S9a) is characterized by a circulating $J^T$ extending over the entire sample with no $J^{NT}$. The induced magnetic monopole can be present only at similar singular points in the parameter space at the vertices of the diamonds at $V_{tg} \cong 0$ and hence is very fragile. Figures S9b2-f2 depict the (0,2) MME state (blue circle in Fig. S9a) characterized by two co-propagating rings of $J^T$ intertwined with counterpropagating $J^{NT}$ currents. In the (-1,1) state (Figs. S9b3-f3 and green circle in Fig. S9a) two counterpropagating rings of $J^T$ are present with adjacent rings of $J^{NT}$. More complex current distributions leading to higher order magnetic moment responses are present in other MME states.

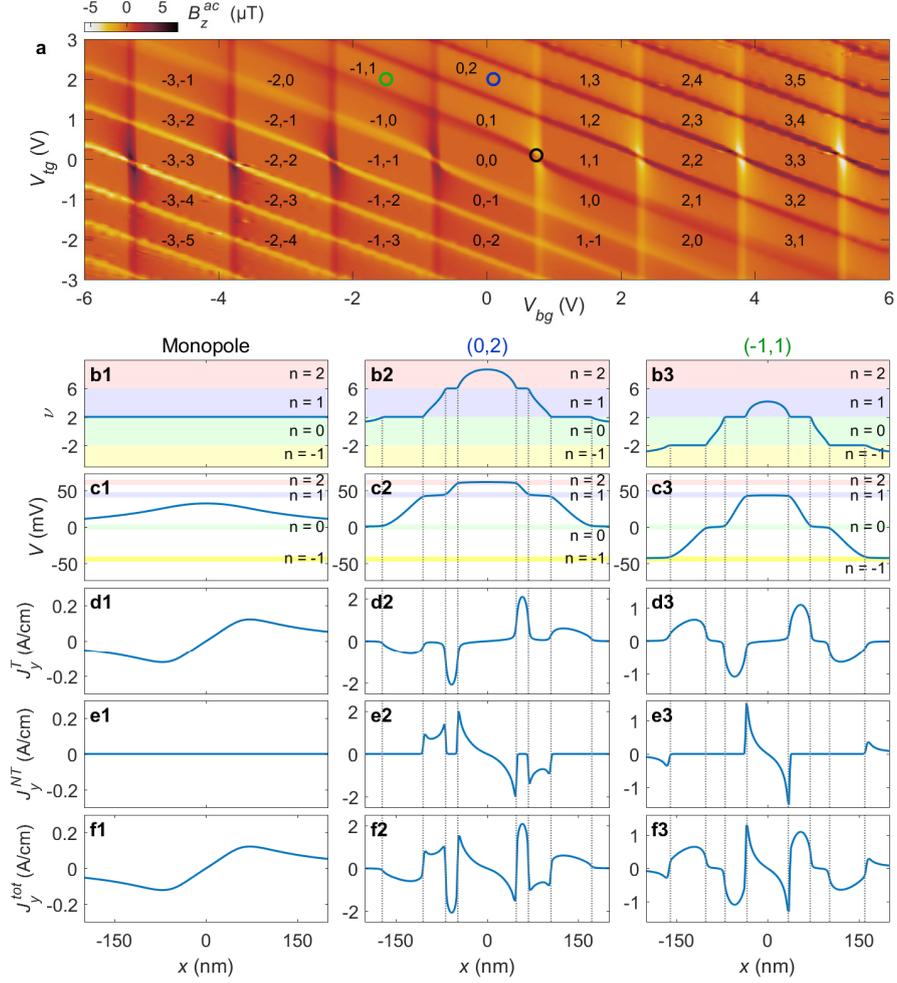

**Fig. S9. Simulation of the phase space of the mixed magnetoelectric effect**. (a) $B_z^{ac}(V_{bg}, V_{tg})$ simulation using the same parameters as Fig. 4c. (**b1-f1**) Calculation of the induced magnetic monopole at $V_{bg} = 0.74$ V and $V_{tg} = 0.1$ V (black circle in (A)) showing the filling factor $\nu$ (**b1**), the potential $V$ (**c1**), the topological current $J^T$ (**d1**) in the incompressible state, the nontopological current $J^{NT}$ (**e1**) in the compressible regions (zero in this case), and the total current (**f1**) vs. position $x$ with the SOT stationed at the origin. (**b2-f2**) same as (b1-f1) showing a condition for a "wedding cake" potential with two co-propagating concentric $J^T$ rings at $V_{bg} = 0.1$ V and $V_{tg} = 2$ V (blue circle in (a)). (**b3-f3**) same as (b1-f1) with conditions for a circular p-n junction with two counter-propagating $J^T$ rings giving rise to the magnetic field of a quadrupole moment at $V_{bg} = -1.5$ V and $V_{tg} = 2$ V (green circle in (a)).

## SM12. Semiclassical numerical simulations

COMSOL simulations were used for solving electrostatic equations for the potential $V$ and charge density $\rho = -en_e$ in graphene. The simulations included one or two metallic backgates, graphene, and the SOT acting as a metallic local top gate (Fig. S10) in a $2 \times 2 \times 2$ μm³ volume with boundary conditions of $E_\perp = 0$ on the box surfaces and a constant electric potential on the backgate and SOT surfaces. We performed an iterative self-consistent solution for $V(x, y, z)$ as follows.

Step 1 – $V(x,y,z)$ was calculated by solving $\nabla \cdot \mathbf{E} = \rho/\varepsilon_r\varepsilon_0$ and $\mathbf{E} = -\nabla V$ with given $V_{bg}$, $V_{tg}$ and an initial charge distribution $\rho(x,y) = \rho_0$ on graphene, where $\varepsilon_r$ is the relative permittivity of the material (we took $\varepsilon_r = 4$ for hBN and 3.9 for SiO$_2$) and $\varepsilon_0$ is the permittivity of vacuum.

Step 2 – The solution for $V(x,y,z=0)$ was used as an input to a smoothed LL occupation function $\nu(V)$ (Fig. S10d) defining the local charge density on the graphene, $\rho(x,y) = eB\nu(V)/\phi_0$. The resulting $\rho(x,y)$ was then used as an input to step 1.

These steps were repeated until a self-consistent solution was attained. Once $V(x,y,z=0)$ and $\rho(x,y)$ were found, the topological and nontopological surface currents in graphene were calculated using $\mathbf{J}^T = -\sigma\nabla V$ and $\mathbf{J}^{NT} = \frac{\mu_e(n)}{e}\nabla \times |\rho|\hat{z}$, where $\sigma_{xy}(x,y) = -\sigma_{yx}(x,y) = -\nu(x,y)e^2/h$ and $\sigma_{xx} = \sigma_{yy} = 0$ are the components of the conductivity tensor $\sigma$. The total current distribution $\mathbf{J}^{tot} = \mathbf{J}^T + \mathbf{J}^{NT}$ is then used to derive the induced magnetic field $B_z(x,y,z)$ through the Bio-Savart relation.

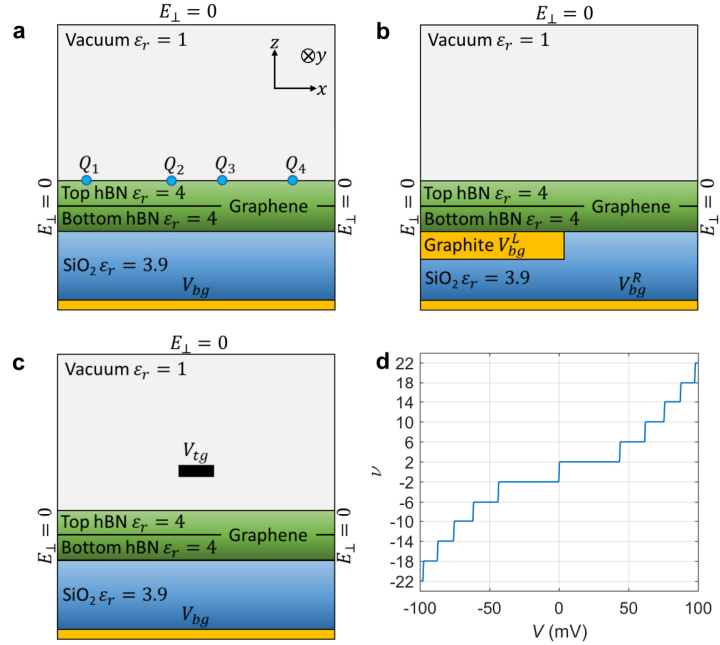

**Fig. S10.** (**a-c**) Electrostatic simulation configurations used for Figs. 2, 3 and 4 respectively, showing the different backgates, SOT top gate, dielectric constants and boundary conditions. (**d**) The function $\nu(V)$ used for electrostatic simulations encoding graphene LLs DOS at $B = 1.45$ T.

### SM13. Microscopic quantum simulations

Additional insight was gained by performing microscopic quantum mechanical simulations. We considered two geometries – linear and cylindrical, corresponding to the situations in Fig. 3 and Fig. 4 respectively. Using the COMSOL calculated electrostatic potential $V(\mathbf{r})$ (section SM12) we applied the Dirac equation in the presence of a uniform magnetic field

$$[v_F \boldsymbol{\sigma} \cdot \mathbf{q} + V(\mathbf{r})]\Psi(\mathbf{r}) = \epsilon\,\Psi(\mathbf{r}),$$

where $\mathbf{q} = -i\hbar\nabla - e\mathbf{A}$ is the kinematic momentum and $q_z = 0$. The vector potential $\mathbf{A}$ was chosen to respect the symmetries of the geometry. Once the eigenstates $\Psi(\mathbf{r})$ were determined, the current density was obtained by

$$J(r) = 4ev_F \sum_\alpha \Psi(r)^\dagger \, \boldsymbol{\sigma} \, \Psi(r),$$

where the summation over $\alpha$ extends over all occupied states assuming zero temperature and the factor of 4 accounts for graphene's spin and valley degeneracies.

The details of the discretization and specific form of the Hamiltonian for the cylindrical geometry are presented in Ref. [31]. The problem of a linear geometry is almost identical to the cylindrical one with the exception of the wave function's ansatz. In the cylindrical geometry, we use the polar decomposition ansatz ($m$ is the angular momentum and $r, \varphi$ are the radial and angular coordinates)

$$\Psi_m(r,\varphi) = \frac{e^{im\varphi}}{\sqrt{r}} \begin{pmatrix} u_1(r) e^{\frac{-i\varphi}{2}} \\ iu_2(r) e^{\frac{i\varphi}{2}} \end{pmatrix},$$

whilst in the linear geometry with translation invariance along $y$ the natural ansatz takes the form

$$\Psi_k(x,y) = e^{-iky} \begin{pmatrix} u_1(x) \\ iu_2(x) \end{pmatrix}.$$

Periodic boundary conditions along $y$ quantize the momentum $k$ according to

$$k_n = \frac{2\pi}{W} n, \, n = \pm 0, 1, 2, \ldots$$

Here, $W$ was numerically taken to be larger than any spatial scale present in $V(x)$.

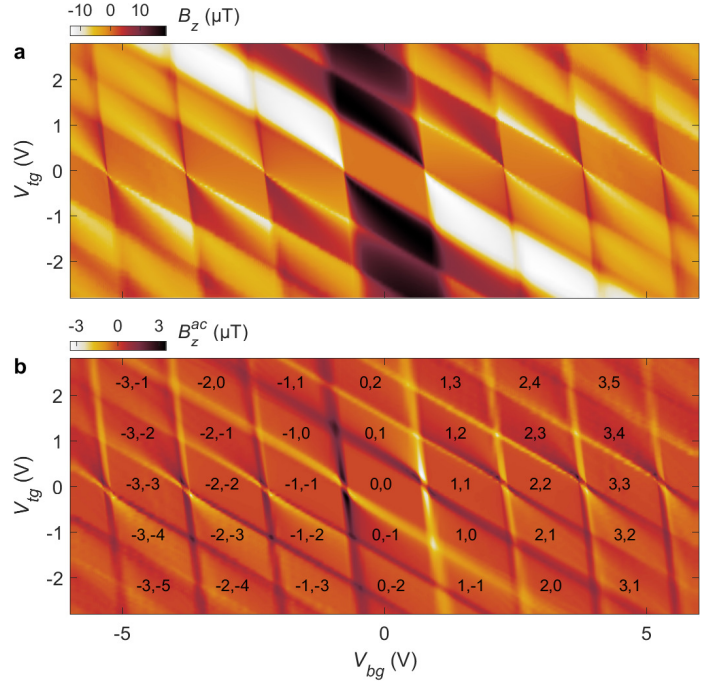

**Fig. S11. Quantum mechanical simulations of the mixed magnetoelectric effect.** Induced $B_z$ (a) and $B_z^{ac}$ (b) at $h = 45$ nm above the graphene for $V_{bg}^{ac} = 50$ mV rms, showing good qualitative agreement with semiclassical calculations (Figs. 4b and 4c) and experimental data (Fig. 4i). The quantum simulations used the same electric potential $V(x,y)$ as in Figs. 4b-c.

The cylindrical simulations were run on a lattice consisting of $N = 480$ sites for a system size $L \approx 320$ nm. The linear geometry used a lattice of $N = 518$, system size $L \approx 500$ nm and $W = 800$ nm. In both cases level broadening $\gamma$ was chosen to be $\approx 2$ meV. As in Ref. [31], the contribution of spurious states present due to a finite system size was excluded.

Fig. 4h compares semiclassical and quantum mechanical calculations for $J^{tot}$ in the circular geometry. The sharp semiclassical features become smeared in the quantum simulation on a length scale of the wave function's width – a few magnetic lengths $l_B$. Adjacent semiclassical $J^T$ and $J^{NT}$ partially cancel out. Nonetheless the overall features can be traced and attributed to regions corresponding to non-zero gradients of potential and of carrier density as argued earlier.

After $\boldsymbol{J}(\boldsymbol{r})$ has been determined, the induced magnetic field $B$ was calculated using the Biot-Savart law. Figures S11a and S11b show the quantum calculation results for $B_z(V_{bg}, V_{tg})$ and $B_z^{ac}(V_{bg}, V_{tg})$. They are in good agreement with the semi-classical derivation in Figs. 4b and 4c.

The *p-n* junction was simulated using a linear geometry with translation invariance along the $y$-direction. Figure S12b displays the resulting $J_y^{QM}(x)$. It is in good qualitative agreement with the semiclassical $J_y^{SC}(x)$ in Fig. S12a. A direct comparison is shown in Fig. S12c for the filling factor marked by the dashed lines in panels A and B. The peaks in $J_y^{QM}(x)$ appear in regions with non- zero gradients in density or potential, but they are smoothened out on the lengthscale of the wave function width of a few $l_B$ as compared to $J_y^{SC}(x)$.

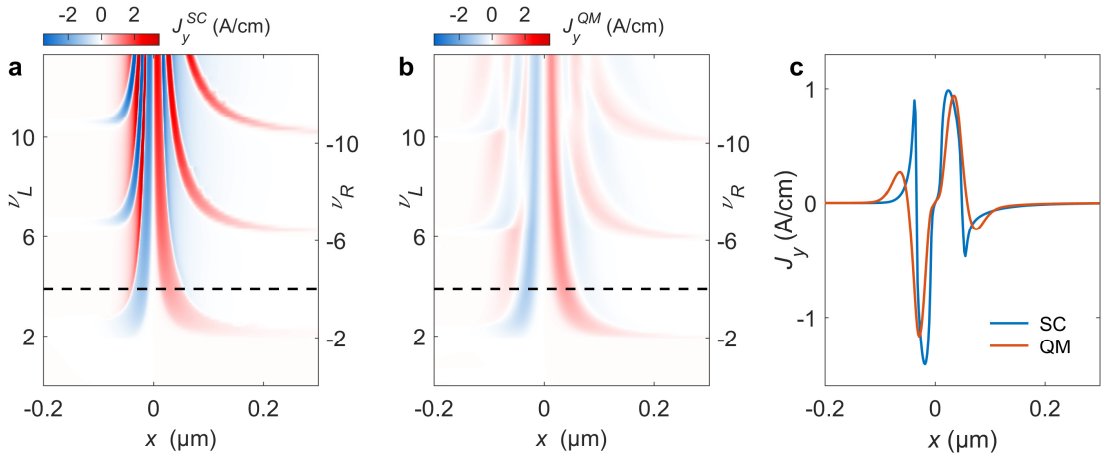

**Fig. S12**. **Comparison of semiclassical and quantum mechanical simulations for the *p-n* junction geometry**. Semiclassical $J_y^{SC}(x)$ (**a**) and quantum mechanical $J_y^{QM}(x)$ (**b**) across a *p-n* junction vs. $\nu_L$ and $\nu_R$ (same setup as Fig. 3). (**c**) $J_y^{SC}(x)$ (blue) and $J_y^{QM}(x)$ (orange) at $\nu_L = 4$ (dashed lines in (a) and (b)) showing that the sharp local features in $J^{SC}$ are smoothened out in $J^{QM}$ on the scale of the wave function of a few magnetic lengths $l_B = \sqrt{\hbar/e|B|}$. Here, $B = 1.45$ T and $l_B \approx 22$ nm.

**SM14. Measurement and simulation parameters**

Figs. 2b-d simulation: $B = 1.435$ T, Si backgate distance from graphene 252 nm, bottom hBN thickness 37 nm, top hBN thickness 15 nm. Other parameters are included in the figure caption.

Fig. 2j simulation: Same as for Figs. 2b-d and with SOT diameter 60 nm, scan height 35 nm (50 nm above graphene), and $x_{ac} = 35$ nm rms.

Fig. 2h: Device A, $B = 1.03$ T, SOT diameter 55 nm, scan height 70 nm, $x_{ac} = 130$ nm rms, pixel size 55 nm, 100 ms per pixel.

Fig. 2i: Device A, $B = 1.05$ T, SOT diameter 55 nm, scan height 60 nm, $x_{ac} = 90$ nm rms, pixel size 95 nm, 60 ms per pixel.

Figs. 3b-e and movie M1: Device B, $B = 1.06$ T, SOT diameter 50 nm, scan height 30 nm, $x_{ac} = 35$ nm rms, pixel size 18 nm, 80 ms per pixel.

Fig. 3f: Device B, $B = 1.06$ T, SOT diameter 50 nm, scan height 25 nm, $x_{ac} = 35$ nm rms, pixel size 13 nm, 80 ms per pixel.

Figs. 3g-j simulation: $B = 1.45$ T, SOT height 30 nm (32 nm above graphene), $x_{ac} = 35$ nm rms, Si backgate distance from graphene 250 nm, graphite backgate distance from graphene 35 nm, top hBN thickness 2 nm.

Figs. 4a-b simulation: $B = 1.415$ T, bottom hBN thickness 17 nm, $SiO_2$ thickness 218 nm, top hBN thickness 15 nm, SOT height 30 nm (45 nm above graphene), backgate excitation (Fig. 4b) $V_{bg}^{ac} = 50$ mV rms.

Fig. 4i: Device A, $B = 1.03$ T, SOT diameter 55 nm, scan height 30 nm, backgate excitation $V_{bg}^{ac} = 70$ mVp square wave at 5.15 kHz, 100 ms per point.

Fig. S6: Device B, $B = 1.06$ T, SOT diameter 50 nm, scan height 40 nm, $x_{ac} = 30$ nm rms, pixel size 60 nm, 60 ms per pixel.

Figs. S7b-e and movie M2: Device C, $B = 0.94$ T, SOT diameter 69 nm, scan height 30 nm, $x_{ac} = 25$ nm rms, pixel size 20 nm, 80 ms per pixel.

Fig. S7f: Device C, $B = 0.94$ T, SOT diameter 69 nm, scan height 30 nm, $x_{ac} = 55$ nm rms, pixel size 10 nm, 60 ms per pixel.